\documentclass[structabstract]{aa}
\usepackage[utf8]{inputenc}
\usepackage{graphicx}
\usepackage{natbib}
\usepackage{xcolor}
\usepackage{hyperref}
\usepackage{widetext}
\usepackage{txfonts}
\usepackage{threeparttable}
\usepackage{lipsum}

\title{An advanced multipole model for (216) Kleopatra triple system
\thanks{Based on observations made with ESO Telescopes
at the La Silla Paranal Observatory under program 199.C-0074 (PI Vernazza).}
}

\titlerunning{An advanced multipole model for (216) Kleopatra triple system}

\author{
    M.~Bro\v{z}\inst{\ref{prague}}         \and 
    F.~Marchis\inst{\ref{seti},\ref{lam}}  \and 
    L.~Jorda\inst{\ref{lam}}               \and 
    J.~Hanu{\v s}\inst{\ref{prague}}       \and 
    P.~Vernazza\inst{\ref{lam}}            \and 
    M.~Ferrais\inst{\ref{lam}}             \and 
    F.~Vachier\inst{\ref{imcce}}           \and 
    N.~Rambaux\inst{\ref{imcce}}           \and 
    M.~Marsset\inst{\ref{mit}}             \and 
    M.~Viikinkoski\inst{\ref{tampere}}     \and 
    E.~Jehin\inst{\ref{liege}}             \and 
    S.~Benseguane\inst{\ref{lam}}          \and 
    E.~Podlewska-Gaca\inst{\ref{poznan}}   \and 
    B.~Carry\inst{\ref{oca}}               \and 
    A.~Drouard\inst{\ref{lam}}             \and 
    S.~Fauvaud\inst{\ref{fauvaud}}         \and 
    M.~Birlan\inst{\ref{imcce},\ref{aira}} \and 
    J.~Berthier\inst{\ref{imcce}}          \and 
    P.~Bartczak\inst{\ref{poznan}}         \and 
    C.~Dumas\inst{\ref{tmt}}               \and 
    G.~Dudzi\'{n}ski\inst{\ref{poznan}}    \and 
    J.~{\v D}urech\inst{\ref{prague}}      \and 
    J.~Castillo-Rogez\inst{\ref{jpl}}      \and 
    F.~Cipriani\inst{\ref{estec}}          \and 
    F.~Colas\inst{\ref{imcce}}             \and 
    R.~Fetick\inst{\ref{lam}}              \and 
    T.~Fusco\inst{\ref{lam},\ref{onera}}   \and 
    J.~Grice\inst{\ref{oca},\ref{ou}}      \and 
    A.~Kryszczynska\inst{\ref{poznan}}     \and 
    P.~Lamy\inst{\ref{lamos}}              \and 
    A.~Marciniak\inst{\ref{poznan}}        \and 
    T.~Michalowski\inst{\ref{poznan}}      \and 
    P.~Michel\inst{\ref{oca}}              \and 
    M.~Pajuelo\inst{\ref{imcce},\ref{puc}} \and 
    T.~Santana-Ros\inst{{\ref{uda},\ref{iccub}}}      \and 
    P.~Tanga\inst{\ref{oca}}               \and 
    A.~Vigan\inst{\ref{lam}}               \and 
    D.~Vokrouhlick\'y\inst{\ref{prague}}   \and 
    O.~Witasse\inst{\ref{estec}}           \and 
    B.~Yang\inst{\ref{eso}}                     
}

   \institute{
     Institute of Astronomy, Faculty of Mathematics and Physics, Charles University, V~Hole{\v s}ovi{\v c}k{\'a}ch 2, 18000 Prague, Czech Republic
     \label{prague}
     \and 
     SETI Institute, Carl Sagan Center, 189 Bernado Avenue, Mountain View CA 94043, USA 
     \label{seti}
     \and 
     Aix Marseille Univ, CNRS, LAM, Laboratoire d'Astrophysique de Marseille, Marseille, France
     \label{lam}
     \and 
     IMCCE, Observatoire de Paris, PSL Research University, CNRS, Sorbonne Universit{\'e}s, UPMC Univ Paris 06, Univ. Lille, France%
     \label{imcce}
     \and 
      Department of Earth, Atmospheric and Planetary Sciences, MIT, 77 Massachusetts Avenue, Cambridge, MA 02139, USA
     \label{mit}
     \and
     Mathematics \& Statistics, Tampere University,  PO Box 553, 33101, Tampere, Finland
     \label{tampere}
     \and 
     Space sciences, Technologies and Astrophysics Research Institute, Universit{\'e} de Li{\`e}ge, All{\'e}e du 6 Ao{\^u}t 17, 4000 Li{\`e}ge, Belgium
     \label{liege}
     \and 
     Faculty of Physics, Astronomical Observatory Institute, Adam Mickiewicz University, ul. S{\l}oneczna 36, 60-286 Pozna{\'n}, Poland
     \label{poznan}
     \and 
     Universit\'e C{\^o}te d'Azur, Observatoire de la C{\^o}te d'Azur, CNRS, Laboratoire Lagrange, France
     \label{oca}
     \and 
     Observatoire du Bois de Bardon, 16110 Taponnat, France 
     \label{fauvaud}
     \and 
     Astronomical Institute of Romanian Academy, 5, Cutitul de Argint Street, 040557 Bucharest, Romania
     \label{aira}
     \and 
     Thirty-Meter-Telescope, 100 West Walnut St, Suite 300, Pasadena, CA 91124, USA
     \label{tmt}
     \and 
     Jet Propulsion Laboratory, California Institute of Technology, 4800 Oak Grove Drive, Pasadena, CA 91109, USA
     \label{jpl}
     \and 
     European Space Agency, ESTEC - Scientific Support Office, Keplerlaan 1, Noordwijk 2200 AG, The Netherlands
     \label{estec}
     \and 
     The French Aerospace Lab BP72, 29 avenue de la Division Leclerc, 92322 Chatillon Cedex, France
     \label{onera}
     \and 
     Open University, School of Physical Sciences, The Open University, MK7 6AA, UK
     \label{ou}
     \and 
     Laboratoire Atmosph\`eres, Milieux et Observations Spatiales, CNRS \& 
     Universit\'e de Versailles Saint-Quentin-en-Yvelines, Guyancourt, France
     \label{lamos}
     \and 
     Secci{\'o}n F{\'i}sica, Departamento de Ciencias, Pontificia Universidad Cat{\'o}lica del Per{\'u}, Apartado 1761, Lima, Per{\'u}
     \label{puc}
     \and 
     Departamento de Fisica, Ingenier\'ia de Sistemas y Teor\'ia de la Señal, Universidad de Alicante, Alicante, Spain
     \label{uda}    
     \and 
     Institut de Ci\'encies del Cosmos (ICCUB), Universitat de Barcelona (IEEC-UB), Martí Franqu\'es 1, E08028 Barcelona, Spain
     \label{iccub}
     \and 
     European Southern Observatory (ESO), Alonso de Cordova 3107, 1900 Casilla Vitacura, Santiago, Chile
     \label{eso}
}

\date{Received x-x-2020 / Accepted x-x-2020}

\abstract
   {}
   {
To interpret adaptive-optics observations of (216) Kleopatra,
we need to describe an evolution
of multiple moons,
orbiting an extremely irregular body
and including their mutual interactions.
Such orbits are generally non-Keplerian
and orbital elements are not constants.
   } 
   {
Consequently, we use a modified $N$-body integrator,
which was significantly extended
to include the multipole expansion of the gravitational field up to the order $\ell = 10$.
Its convergence was verified against the `brute-force' algorithm.
We computed the coefficients $C_{\ell m},S_{\!\ell m}$
for Kleopatra's shape, assuming a~constant bulk density.
For solar-system applications,
it was also necessary to implement a variable distance and geometry of observations.
Our $\chi^2$ metric then accounts for
the absolute astrometry,
the relative astrometry (2nd moon with respect to 1st),
angular velocities,
and also silhouettes,
constraining the pole orientation.
This allowed us to derive the orbital elements of Kleopatra's two moons.
   }
   {
Using both archival astrometric data and
new VLT/SPHERE observations (ESO LP 199.C-0074),
we were able to identify the true periods of the moons,
$P_1 = (1.822359\pm0.004156)\,{\rm d}$,
$P_2 = (2.745820\pm0.004820)\,{\rm d}$.
They orbit very close to the 3:2 mean-motion resonance,
but their osculating eccentricities are too small
compared to other perturbations (multipole, mutual),
so that regular librations of the critical argument are not present.
The resulting mass of Kleopatra,
$m_1 = (1.49\pm0.16)\cdot10^{-12}\,M_\odot$ or 
$2.97\cdot10^{18}\,{\rm kg}$,
is significantly lower than previously thought.
An implication explained in the accompanying paper (Marchis et al.)
is that (216) Kleopatra is a critically rotating body. 
   }
   {}
 

\begin{document}
  \maketitle
\section{Introduction}

(216) Kleopatra was discovered in 1880 by Johann Palisa,
a famous Czech astronomer working at the Austrian observatory located in Croatia
\citep{Palisa_1880AN.....98..129P}.
While we celebrate 140 years of its observational arc,
the time span of observations of moons orbiting Kleopatra
is `only' several tens of years.
It starts from 1980,
when a serendipitous occultation by the outer moon was observed,
or from 2008 \citep{Descamps_etal_2011Icar..211.1022D},
when both moons were discovered using adaptive-optics observations on Keck~II,
till 2019
(this work).
The moons have already been assigned permanent names:
Alexhelios and Cleoselene.

This time span is sufficient to determine not only `static' orbits,
but also analyze their orbital evolution.
In particular, the oblateness of the central body induces nodal precession,
$\dot\Omega = -(3/2)\,nJ_2(R/a)^2\cos i$,
where $J_2$ denotes the zonal quadrupole moment,
$R$~body radius,
$n$~mean motion,
$a$~semimajor axis,
$i$~inclination with respect to the equator
(assuming $e = 0$).
For $J_2 \simeq 0.8$,
it would mean $3\,{\rm deg}\,{\rm d}^{-1}$
for an small-inclination orbit at the distance of 500\,km.
However, (216) Kleopatra is an extreme example.
Its shape is so irregular \citep{Ostro2000,Shepard_etal_2018Icar..311..197S}
that multipoles of higher orders
certainly play some role.
One should use either a~direct integration,
which would be extremely time consuming,
or a~multipole expansion,
as we do in this work.
As an outcome, we determine orbital parameters
with a better accuracy,
by accounting for as many dynamical effects as possible.


\section{Adaptive-optics observations}

For fitting the orbits of Kleopatra moons,
we used three astrometric datasets denoted as
DESCAMPS (from 2008; \citealt{Descamps_etal_2011Icar..211.1022D}),
and SPHERE2017, SPHERE2018,
which were obtained with the VLT/SPHERE instrument \citep{Beuzit_2019A&A...631A.155B}
in the framework of the ESO Large Programme (199.C-0074; PI P.~Vernazza).
A detailed description of all adaptive-optics observations,
their observational circumstances,
reductions, and
resulting astrometric positions
is included in the accompanying paper by Marchis et al.
(see Tabs.~2 and~3 therein),
because most of it shared with the analysis of Kleopatra's shape.

Altogether, the number of measurements is 15 and 18
for the absolute astrometry of the inner and the outer moon, respectively.
For testing purposes, we also used measurements on individual
close-in-time images, which are much more numerous (45 plus 45).
A~conservative estimate of the position uncertainties is
approximately 10\,mas.
We accounted for a systematic shift between the photocentre
and the centre of mass, which is typically a few miliarcseconds.
We used a convex-hull shape model (with zero centre of mass),
rotated and illuminated according to observational circumstances,
and computed its photocentre as the weighted average over all
observable facets in the $(u,v)$ plane. A difference for a non-convex
model is negligible, because the observations were taken close to oppositions.
Alternatively, we used 14 relative astrometry measurements of the two moons,
which partly prevents remaining systematics in the photocentre motion
(or allows their detection).


\section{Moons orbital dynamics}\label{section5}

\newcommand{\xitau}{{\tt Xitau}}

\subsection{N-body model}

For orbital simulations, we use the \xitau\ program%
\footnote{\url{http://sirrah.troja.mff.cuni.cz/~mira/xitau/}},
originally developed for stellar applications
\citep{Broz_2017ApJS..230...19B,Nemravova_etal_2016A&A...594A..55N}.
It is a full N-body model,
based on the Bulirsch-Stoer numerical integrator from the SWIFT package \citep{Levison_Duncan_1994Icar..108...18L},
accounting for mutual interactions of all bodies.
For our purposes, it was necessary to modify it in several ways.
Namely, we implemented:
   (i)~a~fitting of relative astrometry,
  (ii)~angular velocities,
 (iii)~adaptive-optics silhouettes of the primary,
  (iv)~variable distance,
   (v)~variable geometry ($u,v,w$),
  (vi)~brute-force algorithm,
 (vii)~multipole development (up to the order $\ell = 10$; see Section~\ref{sec:multipole}), and
(viii)~external tide (see Section~\ref{sec:tide}).

Consequently, for a comparison of observations of Kleopatra and its moons
with our model, we can use the metric:
\begin{equation}
\chi^2 = w_{\rm sky}\chi^2_{\rm sky} + w_{\rm sky2}\chi^2_{\rm sky2} + w_{\rm sky3}\chi^2_{\rm sky3} + w_{\rm ao}\chi^2_{\rm ao}\,,
\end{equation}
\begin{equation}
\chi^2_{\rm sky} = \sum_{j=1}^{N_{\rm bod}} \sum_{i=1}^{N_{\rm sky}} \left[{(\Delta u_{ji})^2\over\sigma_{{\rm sky\,major}\,ji}^2} + {(\Delta v_{ji})^2\over\sigma_{{\rm sky\,minor}\,ji}^2}\right]\,,
\end{equation}
\begin{equation}
(\Delta u_{ji}, \Delta v_{ji}) = {\bf R}\left(-\phi_{\rm ellipse}-{\pi\over 2}\right) \times \begin{pmatrix}
u'_{ji}-u_{ji} \\
v'_{ji}-v_{ji}
\end{pmatrix}\,,
\end{equation}
\begin{equation}
\chi^2_{\rm sky2} = \sum_{i=1}^{N_{\rm sky2}} \left[{(\Delta u_{i})^2\over\sigma_{{\rm sky\,major}\,i}^2} + {(\Delta v_{i})^2\over\sigma_{{\rm sky\,minor}\,i}^2}\right]\,,
\end{equation}
\begin{equation}
\chi^2_{\rm sky3} = \sum_{i=1}^{N_{\rm sky3}} \left[{(\Delta\dot u_{i})^2\over\sigma_{{\rm sky\,major}\,i}^2} + {(\Delta\dot v_{i})^2\over\sigma_{{\rm sky\,minor}\,i}^2}\right]\,,
\end{equation}
\begin{equation}
\chi^2_{\rm ao} = \sum_{i=1}^{N_{\rm ao}} \sum_{k=1}^{360} {(u'_{ik}-u_{ik})^2 + (v'_{ik}-v_{ik})^2\over\sigma_{{\rm ao}\,i}^2}\,,
\end{equation}
where
the index $i$~corresponds to observational data,
$j$~individual bodies,
$k$~angular steps of silhouette data,
$'$~synthetic data
interpolated to the times of observations~$t_i$
(including the light-time effect).
$u$, $v$ denote the sky-plane coordinates,
$\dot u$, $\dot v$ their temporal derivatives,
${\bf R}$ the rotation matrix,
$\sigma$ observational uncertainties along two axes
(distinguished as 'major', 'minor'),
$\phi_{\rm ellipse}$ angle of the corresponding uncertainty ellipse.
Necessary (216) and Sun ephemerides,
for computations of the variable distance and geometry,
were taken from JPL's Horizons \citep{Giorgini_etal_1996DPS....28.2504G}.

The four terms correspond to
the absolute or 1-centric astrometry (SKY),
relative astrometry (SKY2; i.e. body 3 wrt. 2),
angular velocities (SKY3), and
adaptive-optics silhouettes (AO).
Optionally, we can also use weights,
e.g., $w_{\rm sky3} = 0$,
if the observed $\dot u$, $\dot v$ are systematically underestimated,
or $w_{\rm ao} = 0.3$,
which serves as a regularisation,
preventing unrealistic pole orientations.

Given the overall time span of observations,
our integrations were performed for
$3\,780\,{\rm d}$ (forward) and $1\,{\rm d}$ (backward)
with respect to the epoch $T_0 = 2454728.761806$.
The integrator has an adaptive time step,
with the respective precision parameter
$\epsilon = 10^{-8}$.
The internal time step was typically $0.02\,{\rm d}$,
or smaller if the time was close to the 'time of interest',
i.e., any of the observational data.


\begin{figure}
\centering
\includegraphics[width=8.5cm]{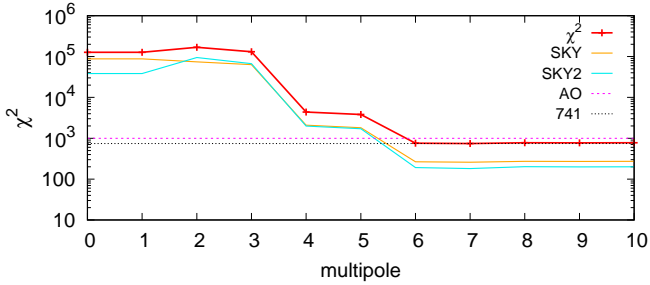}
\caption{
Dependence of $\chi^2 = \chi^2_{\rm sky} + \chi^2_{\rm sky2} + 0.3\,\chi^2_{\rm ao}$
on the multipole order~$\ell$.
The model was optimized for $\ell = 10$
and then recomputed (not optimized) for lower orders.
It is important to account for orders $\ell \le 6$.
}
\label{216_test41_MAXL_chi2_multipole}
\end{figure}

\begin{figure*}
\centering
\begin{tabular}{c@{}c@{}c}
\kern.5cm DESCAMPS &
\kern.5cm SPHERE2017 &
\kern.5cm SPHERE2018 \\
\includegraphics[width=5cm]{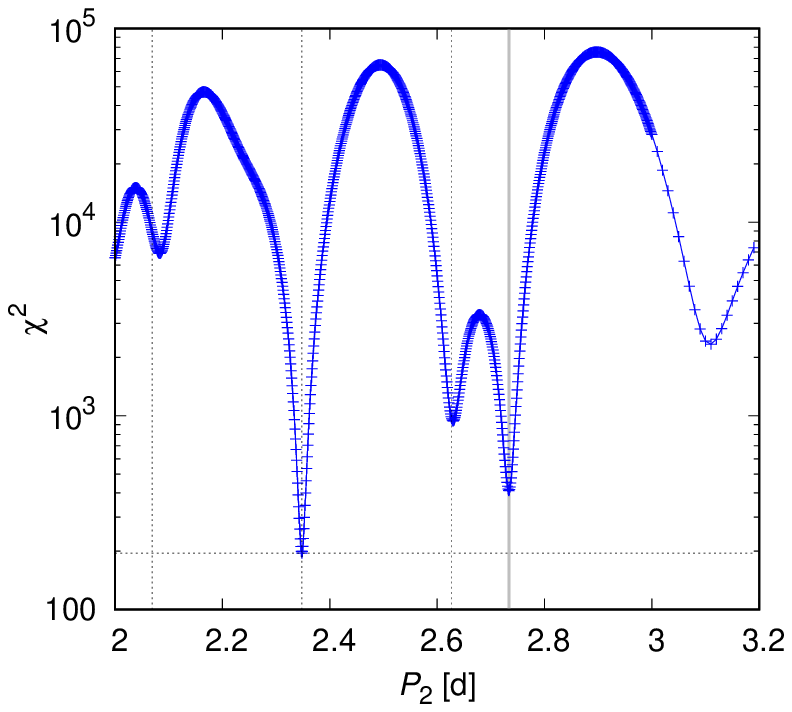} &
\includegraphics[width=5cm]{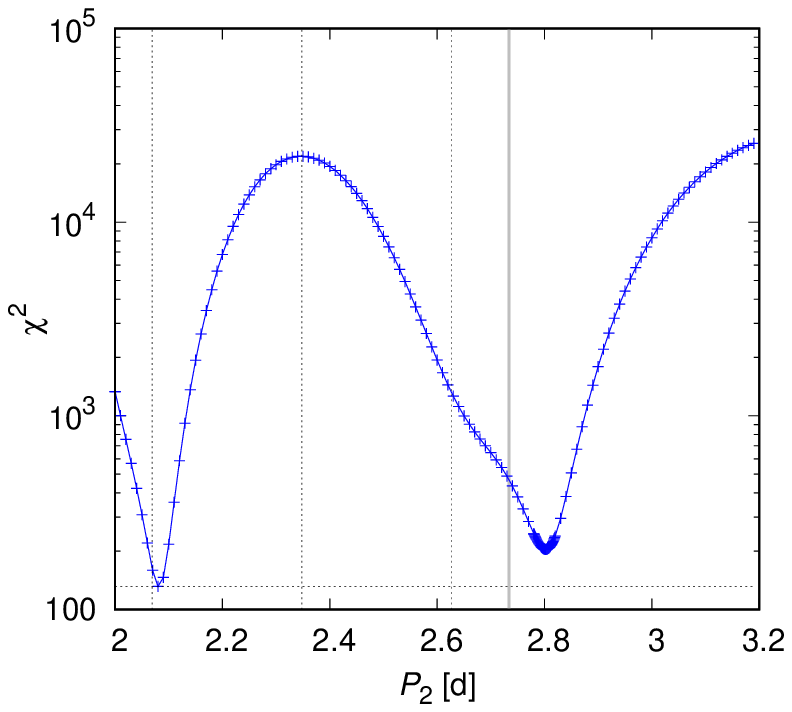} &
\includegraphics[width=5cm]{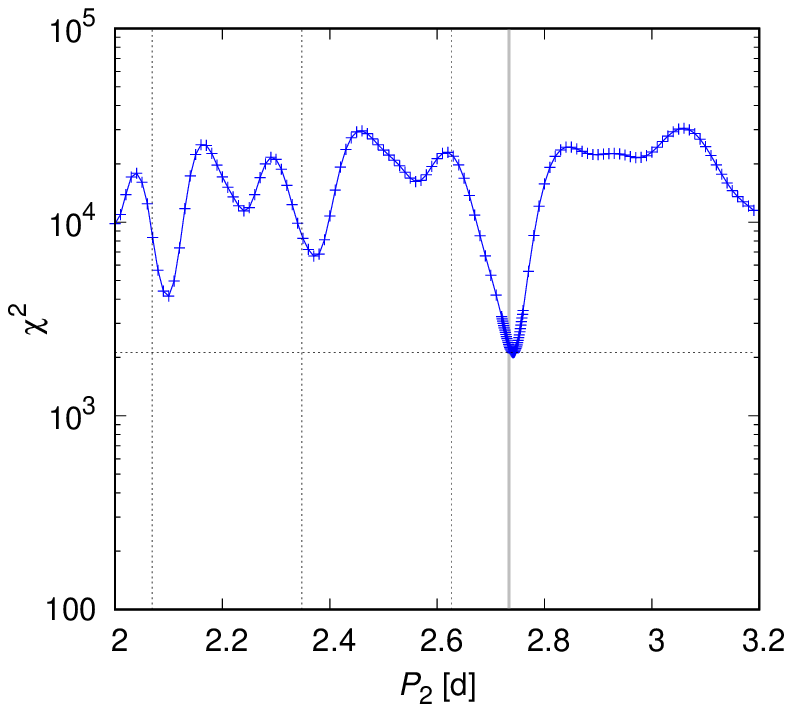} \\
\end{tabular}
\caption{
Periodograms for $P_2$ computed separately for three datasets
(DESCAMPS, SPHERE2017, SPHERE2018).
The $\chi^2 = \chi^2_{\rm sky}$ value was optimized for the first dataset
and then only $P_2$ was varied.
We show the old incorrect period (dotted line),
together with an expected spacing between local minima
given by the timespan $\Delta P = P_2/(t_2-t_1)$,
and the new correct one (gray line).
The shift of $P_2$ for SPHERE2017
and an increased $\chi^2$ for SPHERE2018
were present due to incorrect identification of the two moons.
It was corrected after computing the periodograms and before fitting the orbits.
}
\label{216_fitting2_DESCAMPS_P2}
\end{figure*}

\begin{figure*}
\centering
\begin{tabular}{c@{}c@{}c}
\kern.5cm DESCAMPS &
\kern.5cm SPHERE2017 &
\kern.5cm SPHERE2018 \\
\includegraphics[width=5cm]{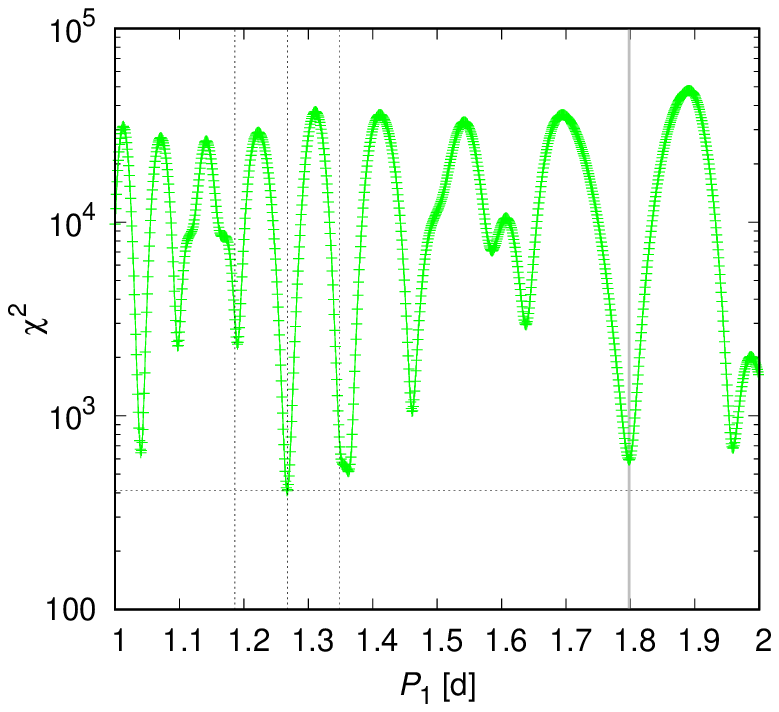} &
\includegraphics[width=5cm]{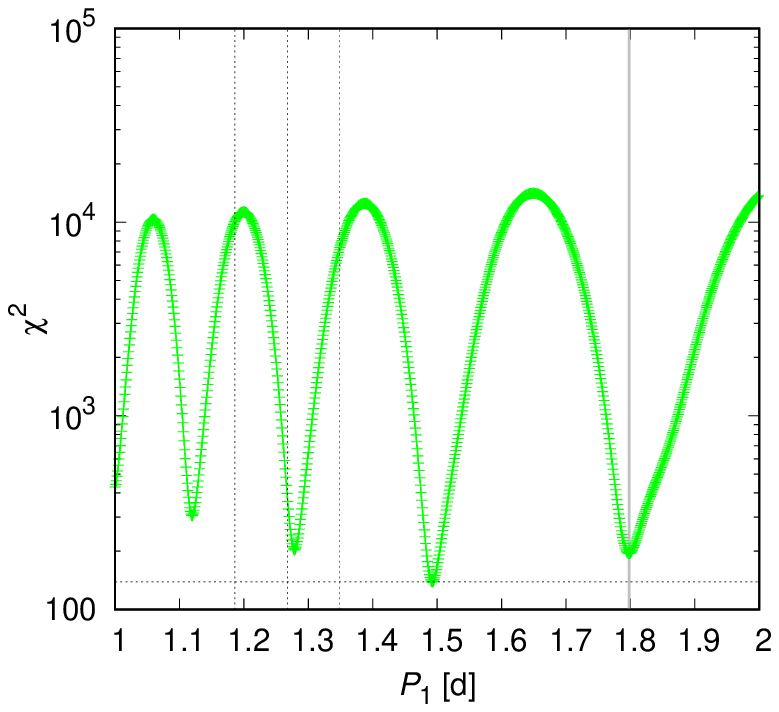} &
\includegraphics[width=5cm]{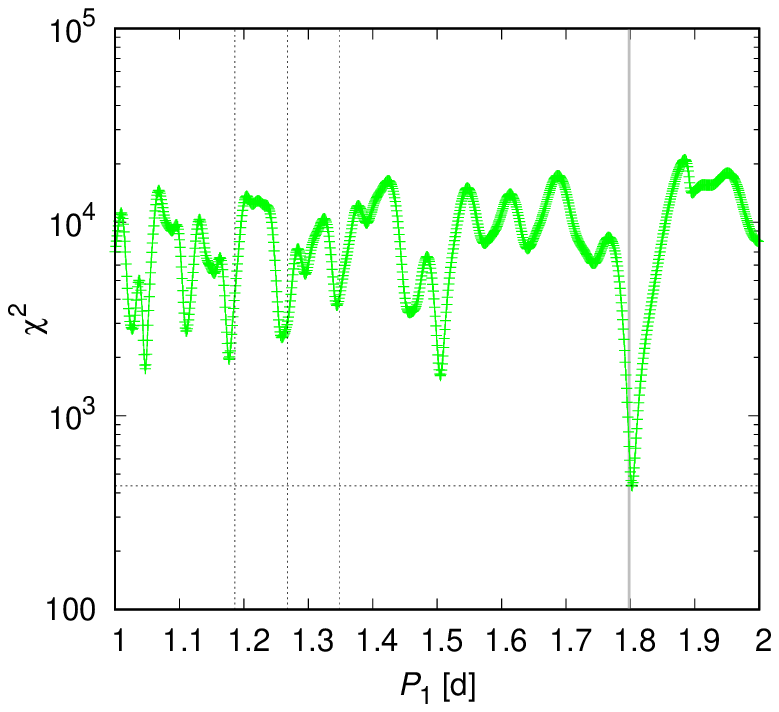} \\
\end{tabular}
\caption{
Same as Fig.~\ref{216_fitting2_DESCAMPS_P2} for $P_1$,
with $P_2$ already shifted towards ${\sim}2.7\,{\rm d}$.
}
\label{216_fitting2_DESCAMPS_P1}
\end{figure*}

\subsection{Brute-force vs. multipole}\label{sec:multipole}

In order to account not only for $J_2$ but for
a gravitational acceleration by an arbitrary shape of the central body,
we implemented a brute-force algorithm in \xitau.
Hereinafter, we assumed a constant density within the body.
The respective volumetric integral:
\def\d{{\rm d}}
\begin{equation}
\vec f_{\rm bf}(\vec r) = -G\rho\int_V {\vec r-\vec r'\over |\vec r-\vec r'|^3}\,{\rm d}V'
\end{equation}
was approximated by a direct sum over 24\,099 tetrahedra,
obtained by a~Delaunay triangulation of the ADAM shape model,
using the Tetgen program \citep{Si_2006}.
The shape was also shifted to the centre of mass
and rotated so that the principal axes of the inertia tensor
correspond to the reference axes.
Although the computation is slow (24\,099 interactions instead of 1),
it can be used as a verification of fast algorithms.

As far as 'fast' is concerned, we also implemented
a multipole development of the gravitational field up to the order $\ell = 10$,
according to \cite{Bursa_etal_1993,Bertotti_etal_2003ASSL..293.....B}.
We review the governing equations here,
using the same notation as in \xitau\ program:

\begin{widetext}
\begin{equation}
U = -{GM\over r}\sum_{\ell=0}^{N_{\rm pole}} \left({R\over r}\right)^{\!\ell} \sum_{m=0}^{\ell} P_{\ell m}(\cos\theta) [C_{\ell m}\cos(m\phi) + S_{\ell m}\sin(m\phi)]\,,\label{eq:U}
\end{equation}
\begin{equation}
{\d U\over\d r} = -GM \sum_{\ell=0}^{N_{\rm pole}} R^\ell (-\ell-1)r^{-\ell-2} \sum_{m=0}^{\ell} P_{\ell m}(\cos\theta) [C_{\ell m}\cos(m\phi) + S_{\ell m}\sin(m\phi)]\,,
\end{equation}
\begin{equation}
{\d U\over\d\theta} = -GM \sum_{\ell=0}^{N_{\rm pole}} R^\ell r^{-\ell-1} \sum_{m=0}^\ell P_{\ell m}'(\cos\theta) \sin\theta [C_{\ell m}\cos(m\phi) + S_{\ell m}\sin(m\phi)]\,,
\end{equation}
\begin{equation}
{\d U\over\d\phi} = -GM \sum_{\ell=0}^{N_{\rm pole}} R^\ell r^{-\ell-1} \sum_{m=0}^\ell P_{\ell m}(\cos\theta) [-C_{\ell m}\sin(m\phi)m + S_{\ell m}\cos(m\phi)m]\,,
\end{equation}
\end{widetext}
\begin{equation}
\vec f_{\rm mp} = -\left({\d U\over\d r}, {1\over r}{\d U\over\d\theta}, {1\over r\sin\theta}{\d U\over\d\phi}\right)\,,
\end{equation}
\begin{equation}
C_{\ell 0} = {1\over MR^\ell} \rho \int_V |\vec r|^\ell P_\ell(\cos\theta) \,{\rm d}V\,,
\end{equation}
\begin{equation}
C_{\ell m} = {2\over MR^\ell} \frac{(\ell-m)!}{(\ell+m)!} \rho \int_V |\vec r|^\ell P_{\ell m}(\cos\theta) \cos(m\phi)\,{\rm d}V\,,
\end{equation}
\begin{equation}
S_{\ell m} = {2\over MR^\ell} \frac{(\ell-m)!}{(\ell+m)!} \rho \int_V |\vec r|^\ell P_{\ell m}(\cos\theta) \sin(m\phi)\,{\rm d}V\,,
\end{equation}
\begin{equation}
P_0(x) = 1\,,\quad P_1(x) = x\,,\quad P_2(x) = {1\over 2}(3x^2-1)\,,\dots
\end{equation}
\begin{equation}
P_{11}(x) = (1-x^2)^{1\over 2}\,,\quad P_{21}(x) = 3x(1-x^2)^{1\over 2}\,,\dots
\end{equation}
where
$r,\theta,\phi$ are body-frozen spherical coordinates
of bodies 2, 3, etc.,
which are determined from 1-centric ecliptic coordinates
by rotations
${\bf R}_z(-l_{\rm pole})$,
${\bf R}_y(-(\pi/2-b_{\rm pole}))$,
${\bf R}_z(-2\pi(t-T_{\rm min})/P - \phi_0)$,
where
$l_{\rm pole}$ denotes the ecliptic longitude of the rotation pole,
$b_{\rm pole}$ ecliptic latitude,
$P$ rotation period,
$T_{\rm min}$ rotation epoch,
$\phi_0$ reference phase;
$R$~the reference radius of the gravitational model,
$U$~gravitational potential,
$\vec f_{\rm mp}$ acceleration,
which is then transformed from spherical to Cartesian and by back-rotations;
$C_{\ell m}$, $S_{\ell m}$ real coefficients,
which have to be evaluated for the given shape model
(see Tab.~\ref{tab1}),
$P_\ell$ the Legendre polynomials, and
$P_{\ell m}$ the associated Legendre polynomials. 
In total, there are 121 dynamical terms in our model.

A verification of convergence is demonstrated in Tab.~\ref{tab2}
(monopole $\to$ brute-force; non-optimized version).
While a difference for the monopole is substantial, $10^{-1}$,
the relative error for $\ell = 10$ is of the order of $10^{-6}$
for the largest $x$-component of acceleration.

Yet the acceleration computation is about 50 times faster (optimized version)
compared to the brute-force algorithm.
An evolution for circular/planar orbits is practically impossible
to distinguish on a 40-day time span; relative differences are of the order
$6\cdot10^{-12}/3\cdot10^{-6} = 2\cdot10^{-6}$.
On the other hand, in extreme cases
(e.g., high inclinations with respect to the equator,
leading to a precession on a $10^2$ day time scale)
there is a noticeable phase shift, resulting in
$4\cdot10^{-8}/3\cdot10^{-6} \simeq 10^{-2}$
variations in ($x, y, z$).

In this work, $C_{\ell m}$, $S_{\ell m}$ coefficients were not
fitted, but kept constant. In principle, it is possible to fit
all of them (with a dedicated version of \xitau),
but it turned out that for almost circular/equatorial orbits
(and sparse astrometric datasets) it is not possible to distinguish
between individual multipoles, which makes the problem degenerate.

In order to understand which multipoles are important,
we estimated $\chi^2$'s for different multipole degrees
(up to some $\ell$, see Fig.~\ref{216_test41_MAXL_chi2_multipole}).
We used an already converged model for $\ell = 10$, without re-convergence, though.
It is clear that the model is very sensitive up to $\ell = 6$.
It may be the case that changing other model parameters (especially $P_1$, $P_2$)
might improve the fits for $\ell < 6$.
Degrees $\ell > 6$ seem to be insignificant for our analysis.


\subsection{External tide}\label{sec:tide}

Additionally, we account for a tide on moons' orbits exerted by the Sun:
\begin{equation}
\vec f_{\rm tidal2} = {GM_\odot\over r_\odot^3} [3(\vec r\cdot\hat n)\hat n - \vec r]\,,
\end{equation}
where 
$M_\odot$ denotes the mass of the Sun,
$r_\odot$ its distance from Kleopatra, and
$\hat n$ its direction with respect to Kleopatra.
It contributes to the satellite orbits precession
by an amount comparable to that from the otherwise included
higher multipole terms of Kleopatra's gravitational field.
We also checked that Jupiter's influence is negligible.

The solar tide also acts on Kleopatra itself.
The related precession of Kleopatra's spin axis is very slow though
and can be neglected in the modelling of its rotation (and shape).
The much faster precession of satellite orbits
(driven by oblateness, or $J_2 \equiv -C_{20}$)
and non-inertial acceleration terms imply that
the Laplace plane always coincides with Kleopatra's equator
\citep{Goldreich_1965AJ.....70....5G},
regardless of any tidal dissipation.


\subsection{Fitting of individual seasons}

Free parameters of our model are as follows:
masses $m_1$,
$m_2$,
$m_3$,
osculating orbital elements of the two orbits
$P_1$,
$\log e_1$,
$i_1$,
$\Omega_1$,
$\varpi_1$,
$\lambda_1$,
$P_2$,
$\log e_2$,
$i_2$,
$\Omega_2$,
$\varpi_2$,
$\lambda_2$
at a given epoch $T_0$,
and the rotation pole orientation
$l_{\rm pole}$,
$b_{\rm pole}$,
i.e., 17 parameters in total.
With \xitau, we can fit any or all of them
with the simplex algorithm \citep{Nelder_Mead_1965}.

Initial values ($P$'s, $m$'s) were taken from \cite{Descamps_etal_2011Icar..211.1022D}.
All $e$'s, $i$'s were "zero" at $t = T_0$, but they are free to evolve.
As a first step, we tried to fit individual datasets.
Regarding DESCAMPS, we immediately reproduced their Fig.~2,
including the suspicious outlier (bottom left).
In fact, it fits on the other side of orbit,
but its error in true longitude is ${\sim}\,90^\circ$!
It is an important observation.

For SPHERE2017 and SPHERE2018, the $\chi^2$ for the nominal $P$'s was
excessively large. It is an indication that the true periods might be
either shorter or longer. Consequently, we computed periodograms
(as $\chi(P)$) for a wide range of periods
(see Figs.~\ref{216_fitting2_DESCAMPS_P2}, \ref{216_fitting2_DESCAMPS_P1}).
It was quite important to start with $P_2$,
because the true period is longer,
and this allowed us to realize that $P_1$ is also longer.
Otherwise, $P_1$, $P_2$ were so close to each other
that the moon system became totally unstable.

After recomputing the periodograms, we obtained preliminary values
of the true periods:
$P_1 = (1.818 \pm 0.010)\,{\rm d}$,
$P_2 = (2.740 \pm 0.010)\,{\rm d}$.
The uncertainties are still large, because seasons have been
treated separately.
Nevertheless, the corresponding mass $m_1$ of Kleopatra should be clearly
much lower than derived in previous works.
It will turn out later that a low~$m_1$ implies
Kleopatra is actually very close to a critical surface,
which we think is {\em not\/} a coincidence.


\subsection{Fitting of DESCAMPS\,+\,SPHERE}

As a next step, we fitted all datasets together.
This required not only a substantially longer time span
($3780\,{\rm d}$ vs $40\,{\rm d}$),
but also a 2-dimensional periodogram with a fine spacing,
$\Delta P \simeq P^2/(t_2-t_1) \simeq 10^{-3}\,{\rm d}$.
We simply cannot use 1-dimensional periodograms for $P_1$ and $P_2$,
because the moons are interacting.
If we change $P_1$ (only), $\chi^2$ for $P_2$ also changes (albeit more slowly).
The only way how to find a joint minimum is to try all combinations.
Given the period uncertainties are at least several $10^{-2}\,{\rm d}$,
this represents about $10^3$ combinations. For each of the (initial) values,
we performed 50 iterations by simplex (with both $P_1$ and $P_2$ free).%
\footnote{1~iteration takes ${\sim}\,10\,{\rm min}$, in total 1~week on 70\,CPUs}
We verified that this was enough to reach a local minimum.
This way, we can be sure that we did not miss a global minimum.
The result is shown in Fig.~\ref{216_fitting9_BIGGRID_P1_P2_min_32}.
It is not a simple $\chi^2$ map --- every point is a local minimum.
Apart from blue areas, there are many local minima in between,
where the simplex is stuck.
Global-minimum algorithms (e.g., simulated annealing, differential evolution, genetic)
are not very useful here, because one would have to try all combinations anyway.

Now, we can re-iterate the problem: we want to make all parameters free, but if we
change `anything' in our dynamical model, then we may be offset from our previously
found local minimum of $P_1$, $P_2$. We also have to check neighbouring local
minima! In other words, some perturbations (e.g., the precession of $\Omega$, $\varpi$)
can be compensated for by an adjustment of $P_1$, $P_2$. This is especially
true for almost circular and almost equatorial orbits, where we cannot recognise
the precession or $e > 0$, $i > 0$ in sky-plane motions, only as a phase shift.

Consequently, we iterated parameters sequentially,
with help of several finer grids (in $P_1$, $P_2$).
We also re-measured one outlier
and included the relative astrometry (SKY2)
in order to check for possible systematics.
In particular, we confirmed that $m_1$ is indeed low,
around $1.5\cdot10^{-12}\,M_\odot$,
with the corresponding bulk density
$\rho_1 = 3\,300\,{\rm kg}\,{\rm m}^{-3}$.
The minimum reached so far is
$\chi^2 = \chi^2_{\rm sky} + \chi^2_{\rm sky2} = 315$.

\begin{figure}
\centering
\includegraphics[width=8.5cm]{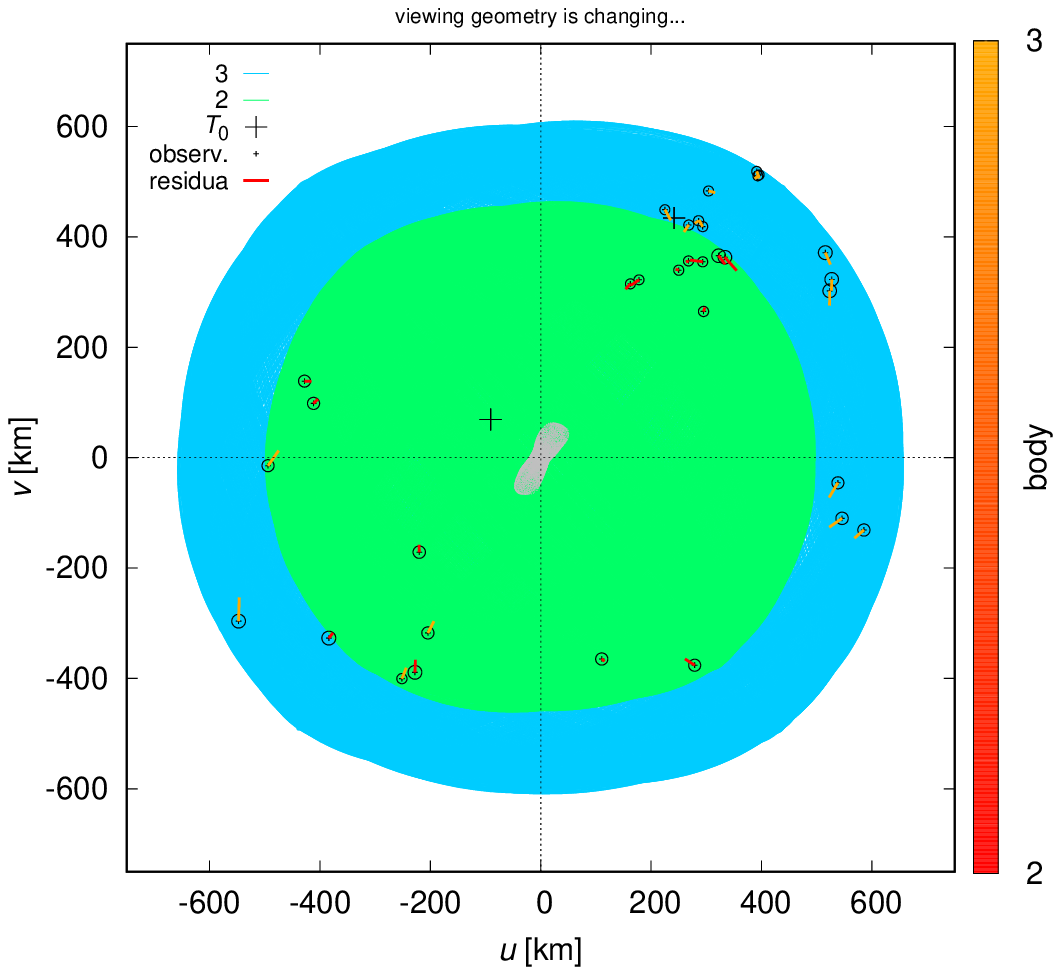}
\hbox{\kern-.4cm\includegraphics[width=7.7cm]{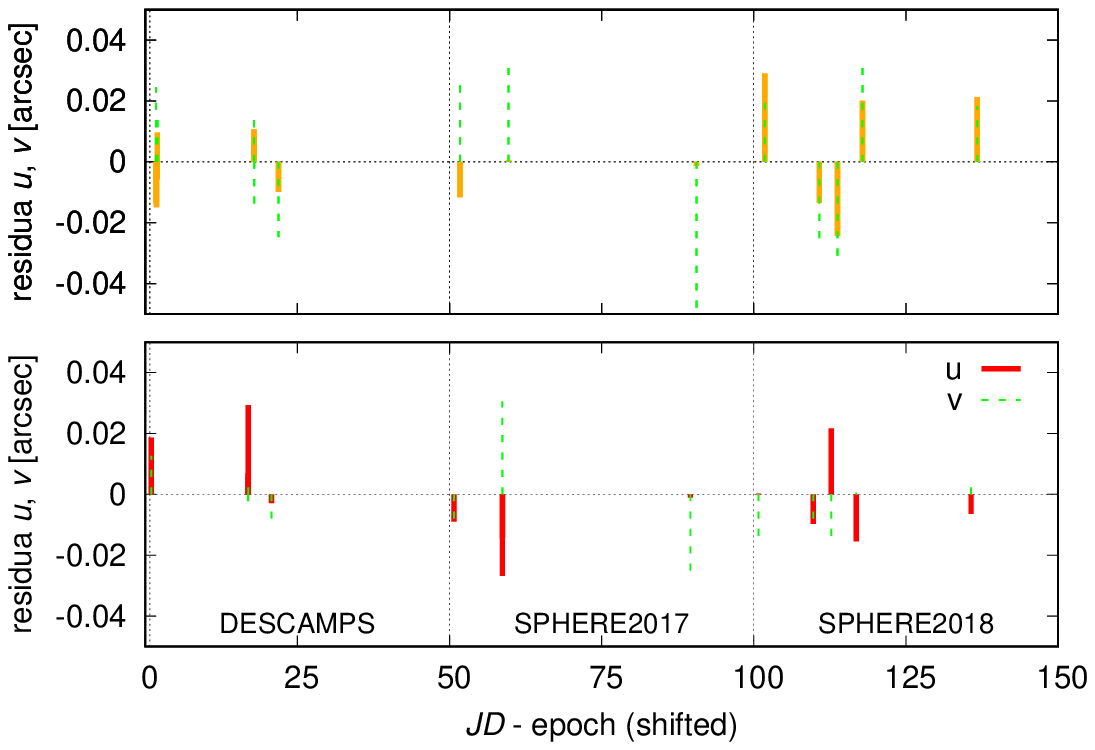}}
\caption{
Best-fit model with $\chi^2 = \chi^2_{\rm sky} + \chi^2_{\rm sky2} + 0.3\,\chi^2_{\rm ao} = 368$.
Top:
The orbits of Kleopatra's moons plotted in the $(u,v)$ coordinates ({\color{blue}blue}, {\color{green}green} lines),
observed absolute astrometry (SKY; black circles),
and residua ({\color{red}red}, {\color{orange}orange} lines
for bodies 2, 3, i.e., inner and outer satellites).
Kleopatra's shape model for one of the epochs is overplotted in gray.
The axes are scaled in km; with a variable viewing geometry,
but without a variable distance.
The mean semimajor axes of orbits are:
$a_1 \doteq 499\,{\rm km}$,
$a_2 \doteq 655\,{\rm km}$.
Bottom:
The residua of $(u,v)$ in arcsec
for the epochs of three datasets (DESCAMPS, SPHERE2017, SPHERE2018).
The uncertainties of astrometric observations were approximately 0.01\,arcsec.
}
\label{216_fitting33_360DEG_chi2_SKY_uv}
\end{figure}

\begin{figure}
\centering
\includegraphics[width=8.5cm]{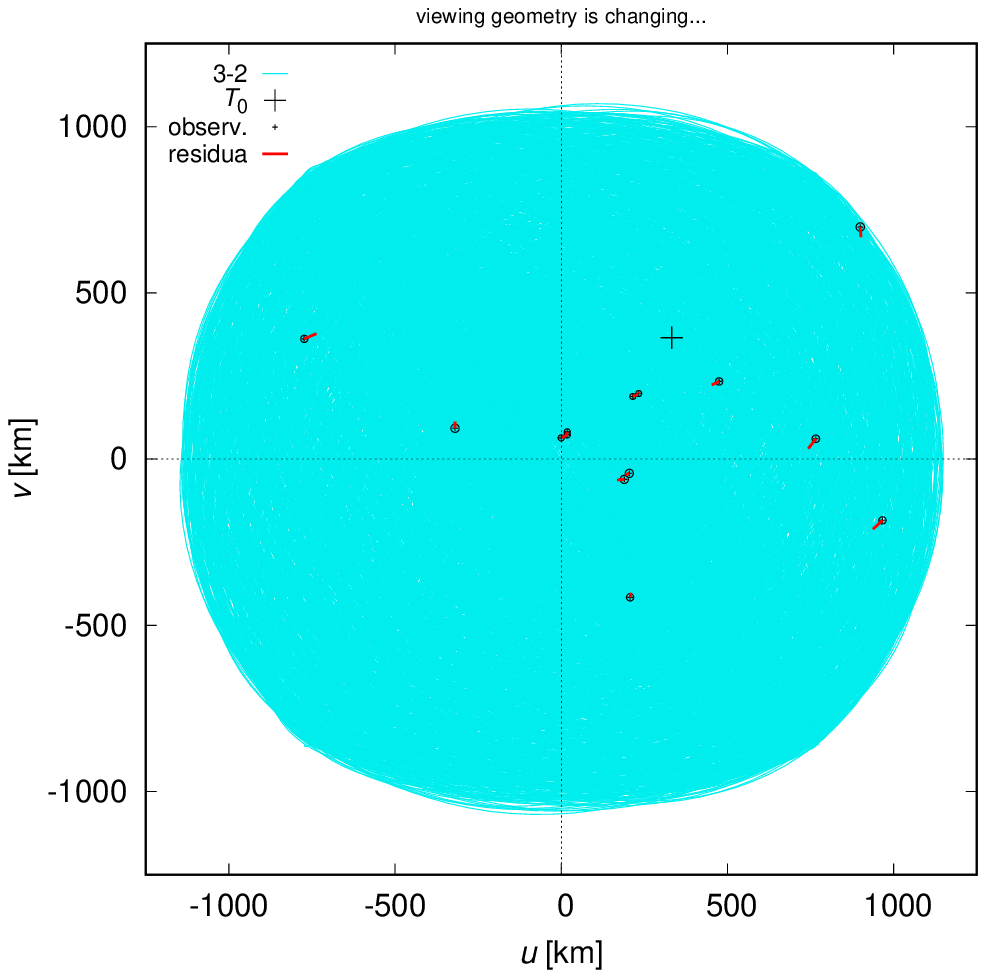}
\hbox{\kern.4cm\includegraphics[width=8.5cm]{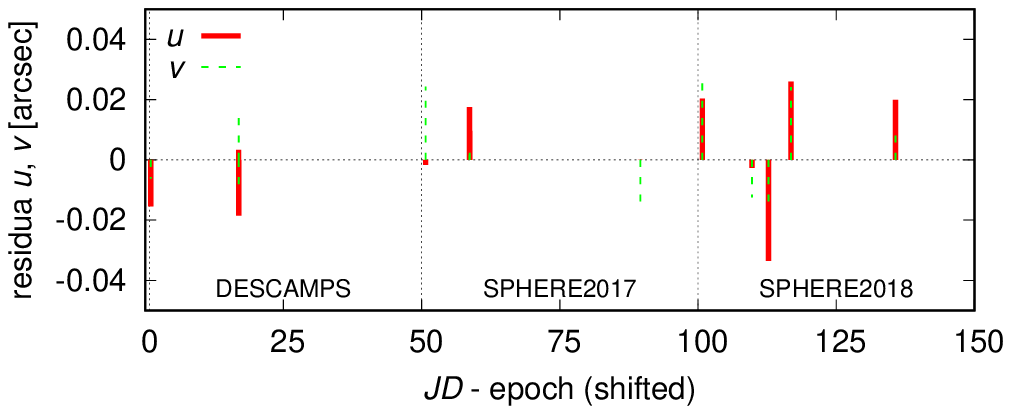}}
\caption{
Same as Fig.~\ref{216_fitting33_360DEG_chi2_SKY_uv},
but for the relative astrometry (SKY2; 3rd body wrt. 2nd).
Point (0,0) is thus centered on the inner moon.
}
\label{216_fitting33_360DEG_chi2_SKY2_uv}
\end{figure}

\begin{figure}
\includegraphics[width=8.5cm]{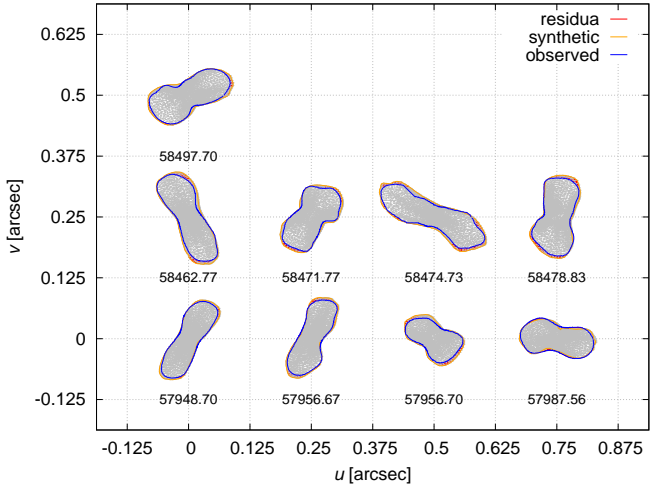}
\caption{
Silhouettes of Kleopatra in $(u,v)$ coordinates ({\color{orange}orange}),
computed for nine epochs (${\rm JD}-2400000.0$),
compared to SPHERE2017, SPHERE2018 observations ({\color{blue}blue}),
and residua ({\color{red}red}).
}
\label{216_fitting33_360DEG_chi2_AO}
\end{figure}

\begin{figure}
\centering
\includegraphics[width=8.5cm]{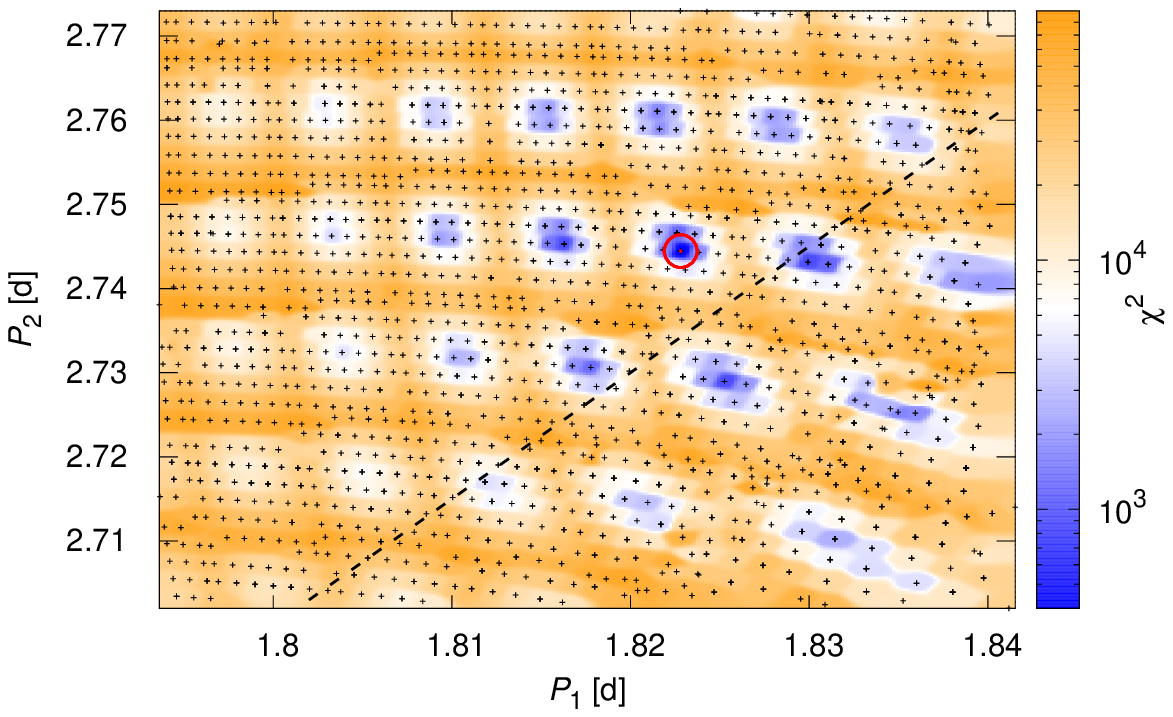}
\caption{
$\chi^2 = \chi^2_{\rm sky}$ values for a range of periods $P_1$, $P_2$
and optimized models. Every black cross denotes a local minimum
(i.e., not a simple $\chi^2$ map).
All datasets (DESCAMPS, SPHERE2017, SPHERE2018) were used together,
consequently, the spacing between local minima is very fine.
The global minimum is denoted by a red circle. 
The dashed line indicates the exact 3:2 period ratio.
}
\label{216_fitting9_BIGGRID_P1_P2_min_32}
\end{figure}


\subsection{Moon masses}

\begin{figure}
\centering
\includegraphics[width=8.0cm]{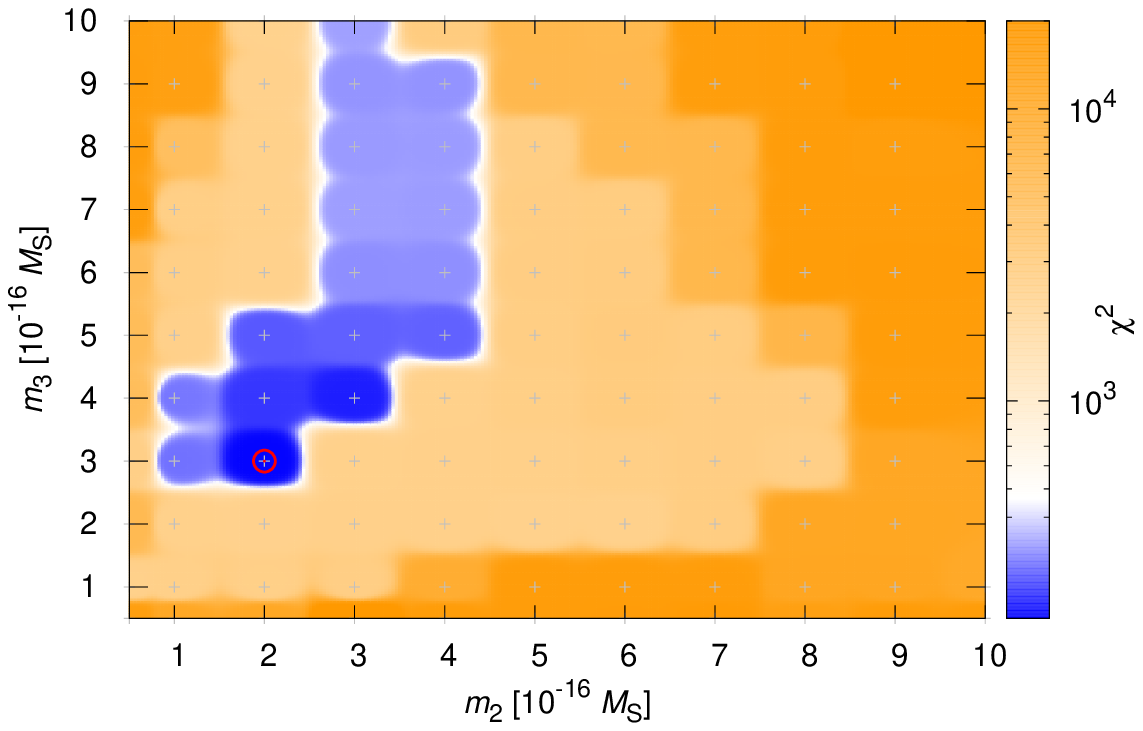}
\caption{
$\chi^2 = \chi^2_{\rm sky} + \chi^2_{\rm sky2}$ values for a range
of moon masses $m_2$, $m_3$.
All models were optimized with respect to the periods $P_1$, $P_2$.
Other parameters were fixed.
The global minimum is denoted by a red circle. 
}
\label{216_fitting23_MOONMASS_m2_m3_min}
\end{figure}

We also looked for the optimum masses of the moons
(Fig.~\ref{216_fitting23_MOONMASS_m2_m3_min}).
It turned out they are around:
$m_2 = 2\cdot10^{-16}\,M_\odot$,
$m_3 = 3\cdot10^{-16}\,M_\odot$,
which together with diameters \citep{Descamps_etal_2011Icar..211.1022D}:
$D_2 = 6.9\,{\rm km}$,
$D_3 = 8.9\,{\rm km}$,
would correspond to the densities:
$\rho_2 = 2\,300\,{\rm kg}\,{\rm m}^{-3}$,
$\rho_3 = 1\,600\,{\rm kg}\,{\rm m}^{-3}$.
These are somewhat lower than Kleopatra's value,
but the 1-$\sigma$ uncertainties are still too large (50\,\%)
to be conclusive.

For example, the case with
$\rho_1 = \rho_2 = \rho_3$
(i.e., $m_2 = 3\cdot10^{-16}\,M_\odot$, $m_3 = 6\cdot10^{-16}\,M_\odot$)
is marginally (3-$\sigma$) allowed, having
$\chi^2 = \chi^2_{\rm sky} + \chi^2_{\rm sky2} = 305$ vs. $182$.
A possibility of massive moons
($\rho_2, \rho_3 > \rho_1$),
especially when we increase
$m_1 = 1.65\cdot10^{-12}\,M_\odot$
at the same time, is also allowed, with
$\chi^2 = 205$ vs. $182$.
A hypothetical possibility
of 'zero-mass' moons, with $\chi^2 = 214$ vs. $182$,
after a manual adjustment of $P_1$, $P_2$,
cannot be excluded, nonetheless,
if we believe in $D{\rm's} > 0$, we should believe in $m{\rm's} > 0$.
Interactions of the moons are inevitable\dots

\subsection{Best-fit and alternative model}

\begin{figure*}
\centering
\includegraphics[width=6cm]{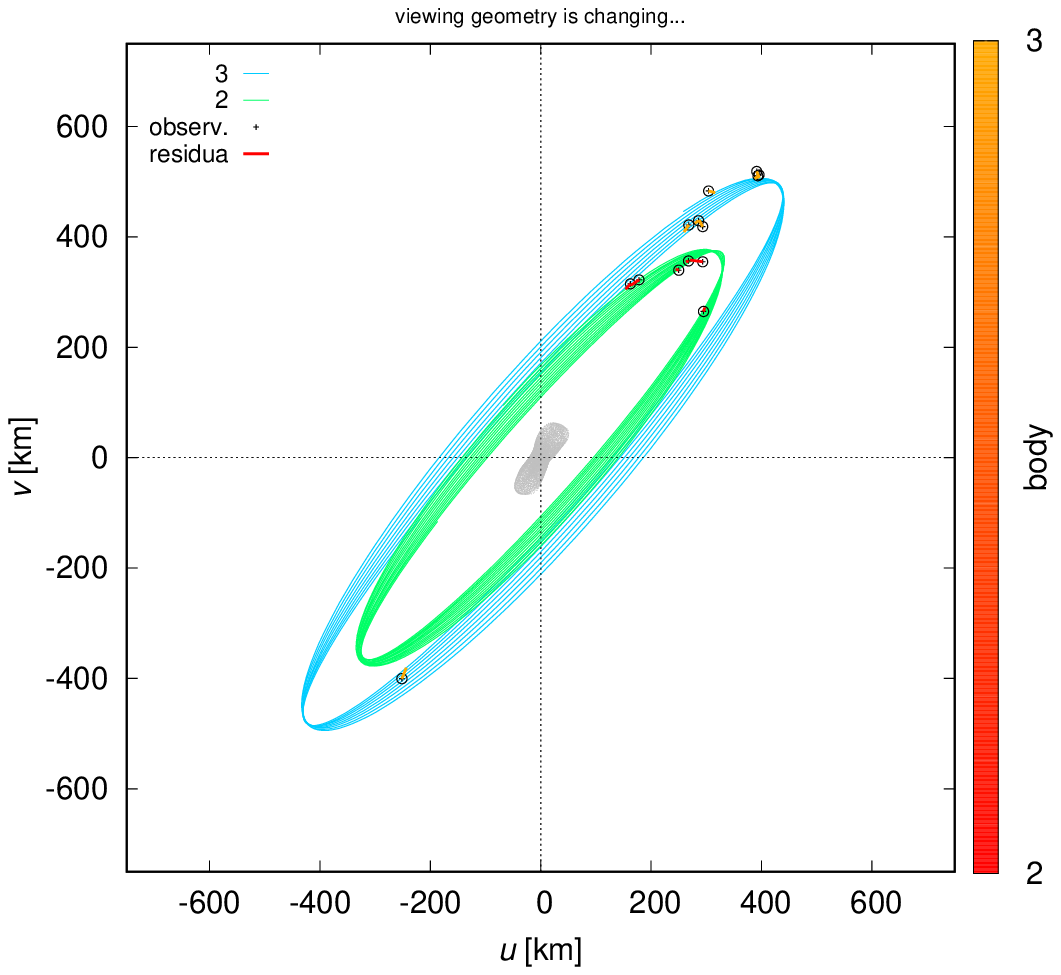}
\includegraphics[width=6cm]{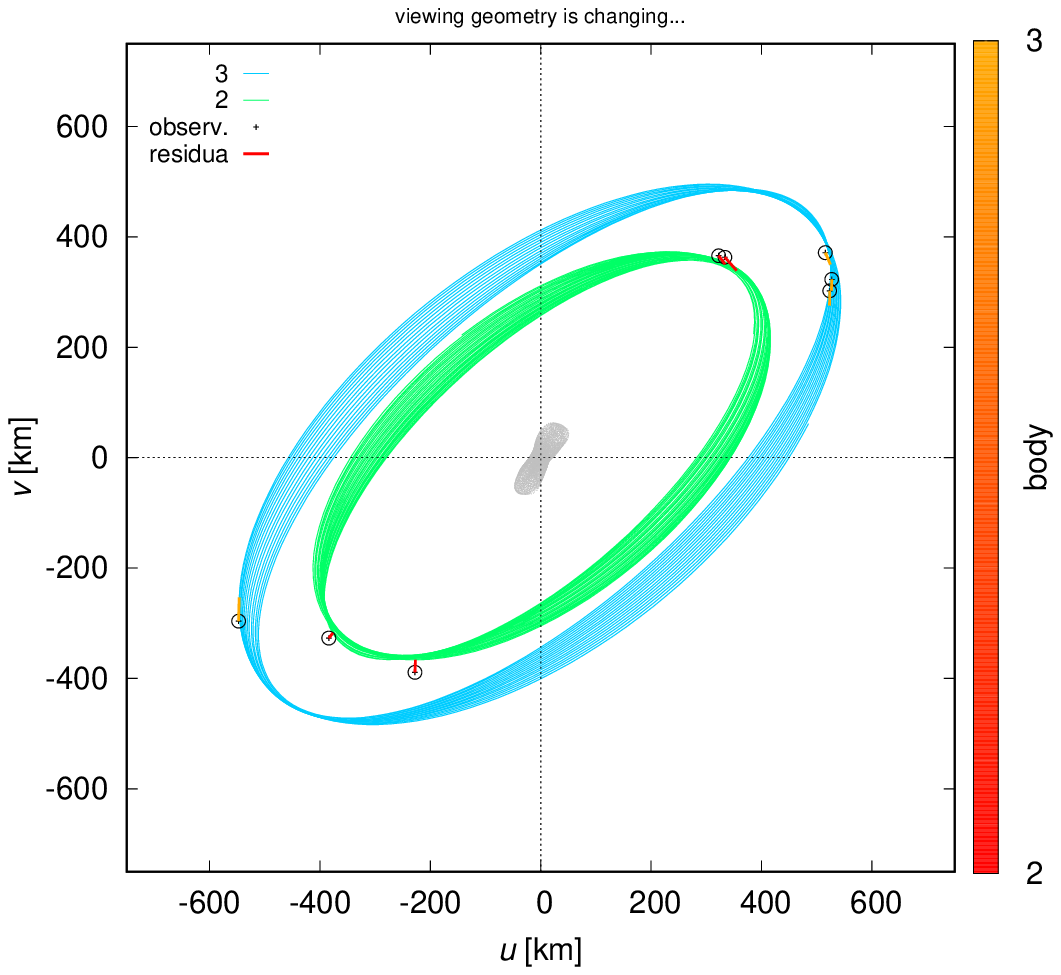}
\includegraphics[width=6cm]{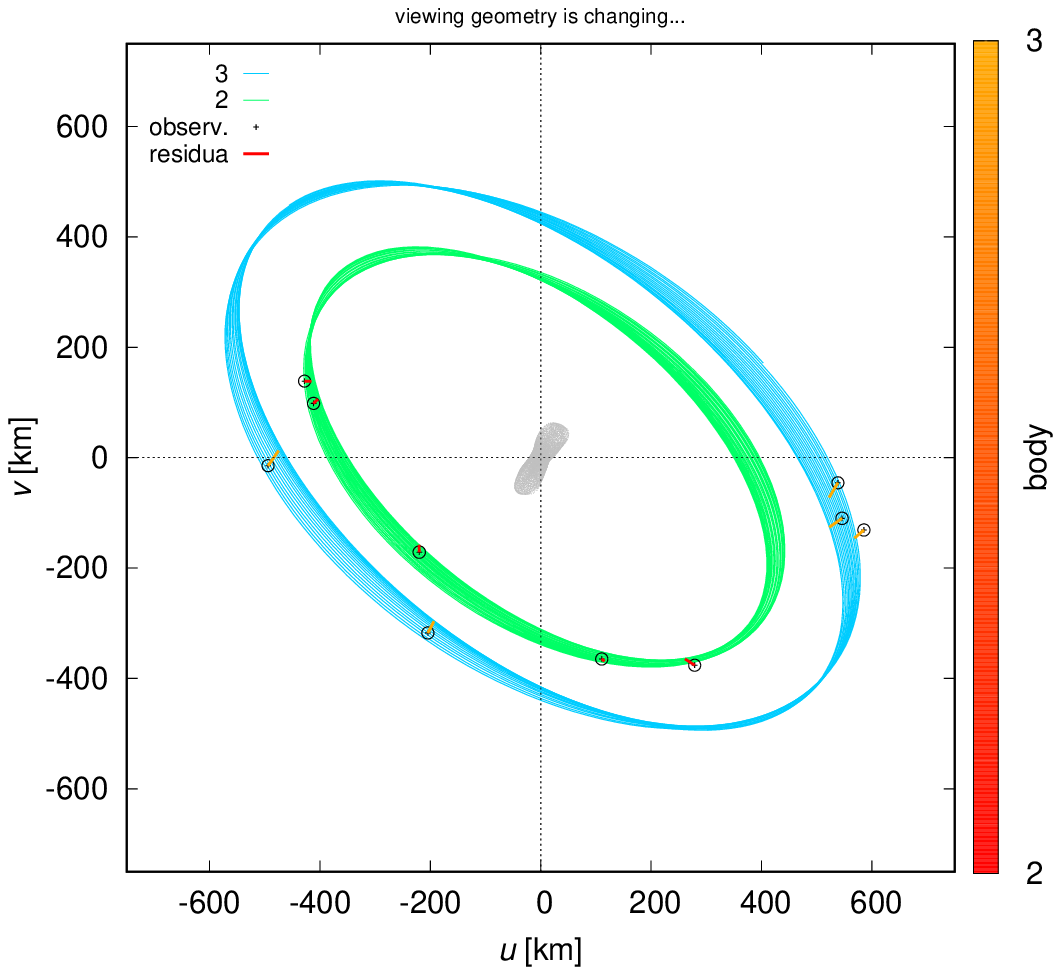}
\caption{
Same as Fig.~\ref{216_fitting33_360DEG_chi2_SKY_uv},
but plotted separately for the three datasets. 
}
\label{216_fitting33_360DEG_DESCAMPS_chi2_SKY_uv}
\end{figure*}

\begin{figure*}
\centering
\includegraphics[width=5.7cm]{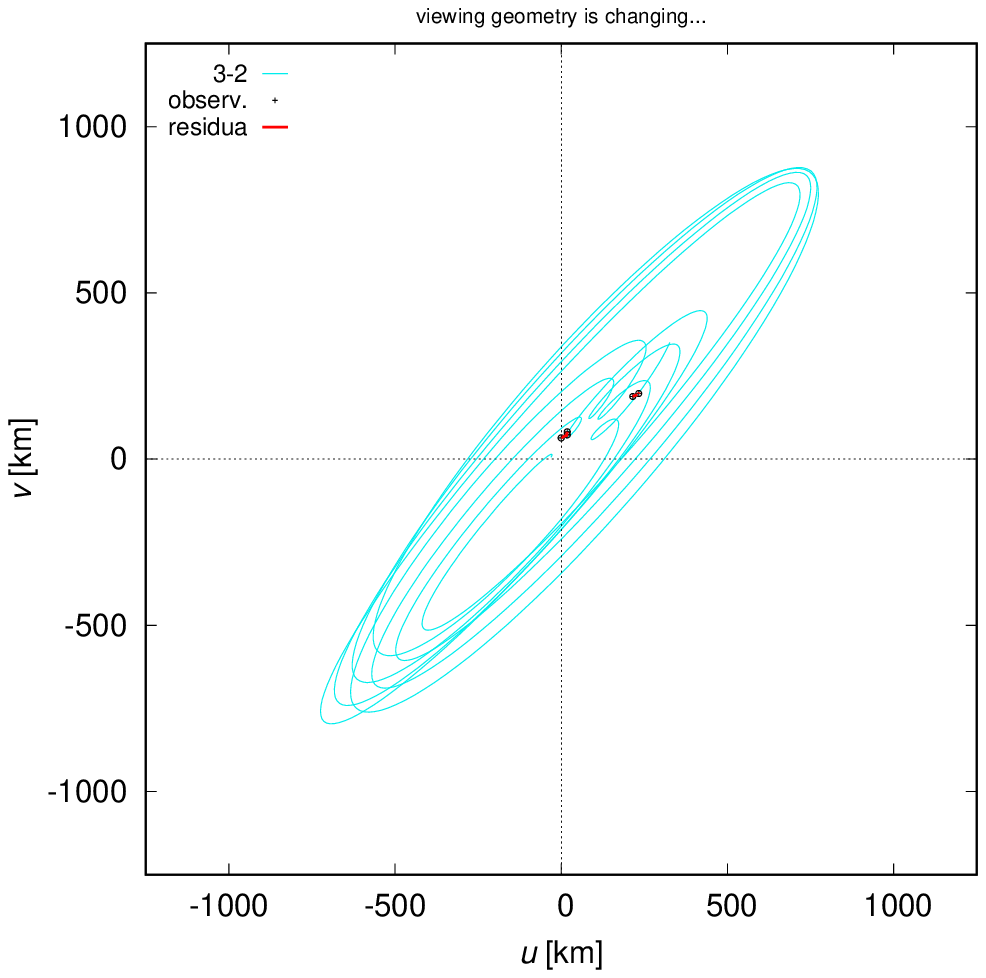}
\includegraphics[width=5.7cm]{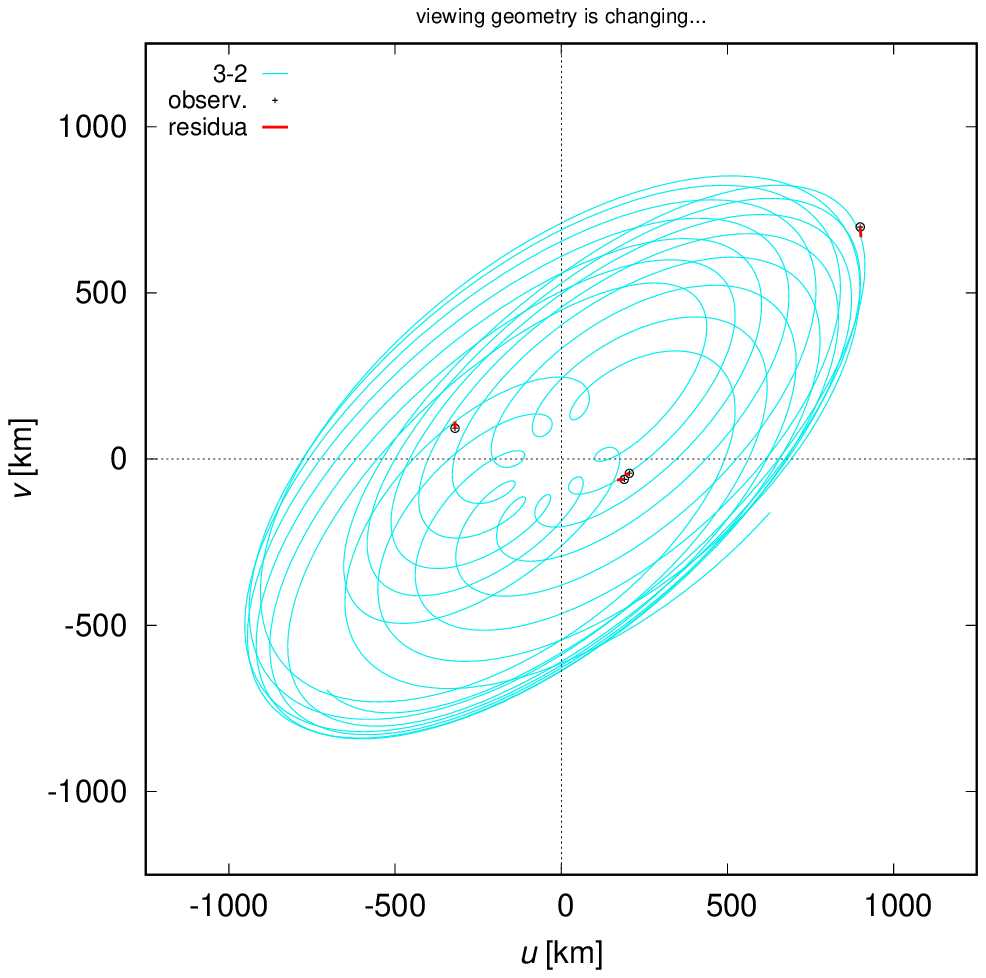}
\includegraphics[width=5.7cm]{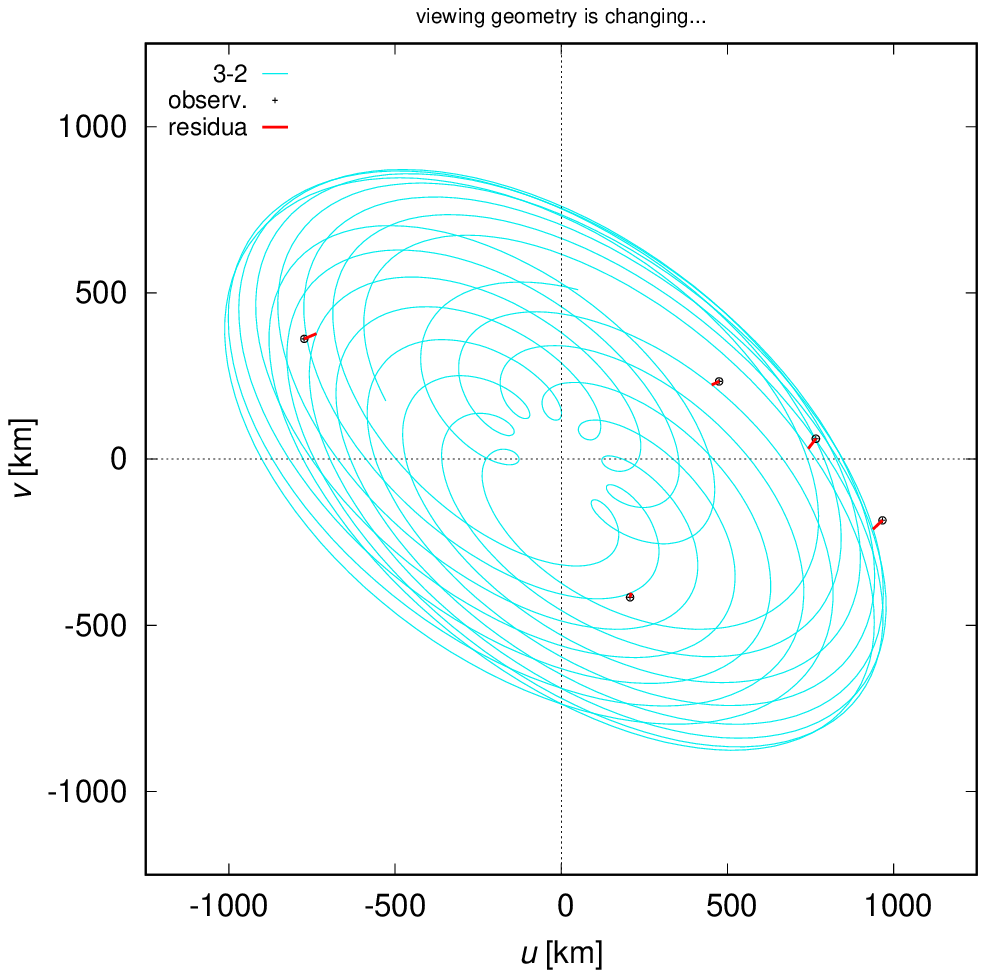}
\caption{
Same as Fig.~\ref{216_fitting33_360DEG_chi2_SKY2_uv},
but plotted separately for the three datasets. 
}
\label{216_fitting33_360DEG_DESCAMPS_chi2_SKY2_uv}
\end{figure*}

\begin{figure*}
\centering
\includegraphics[width=9cm]{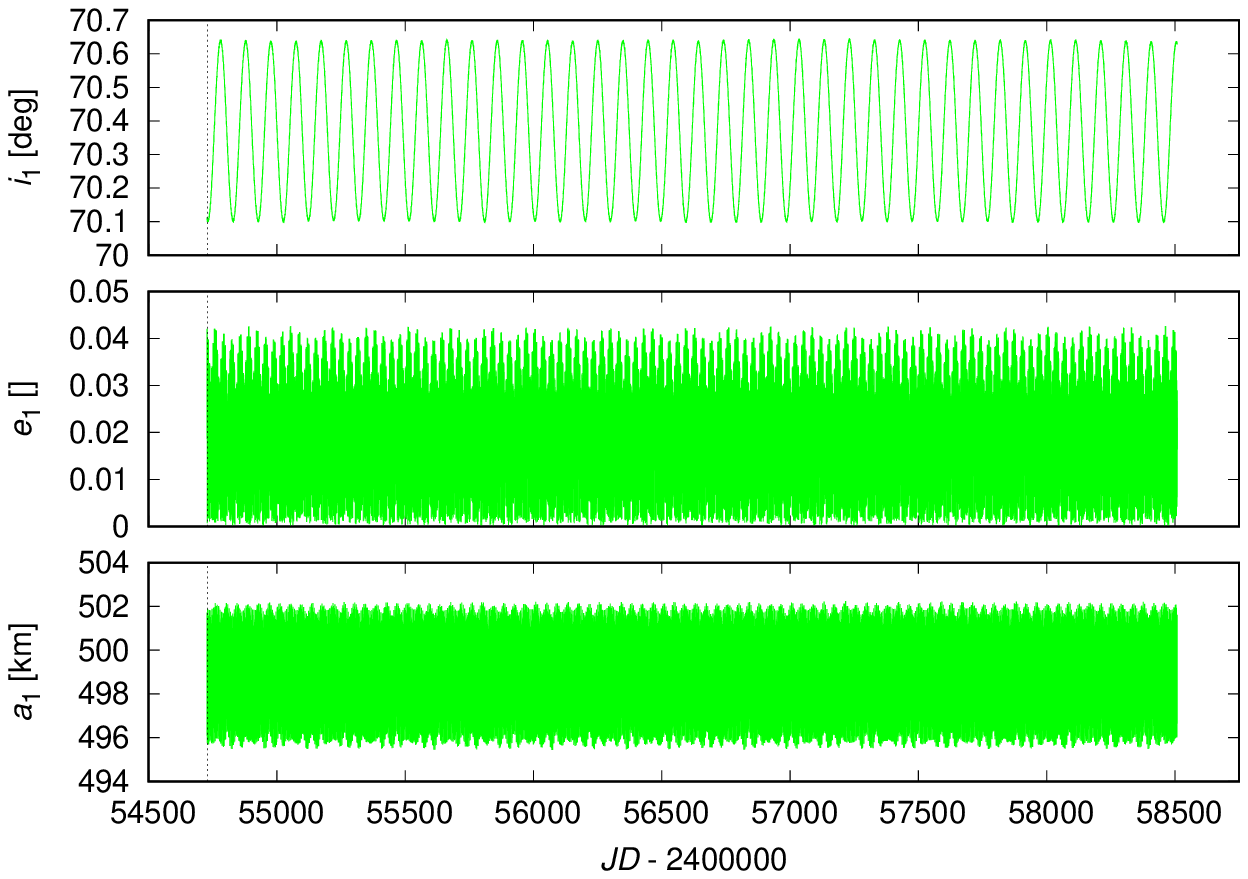}
\includegraphics[width=9cm]{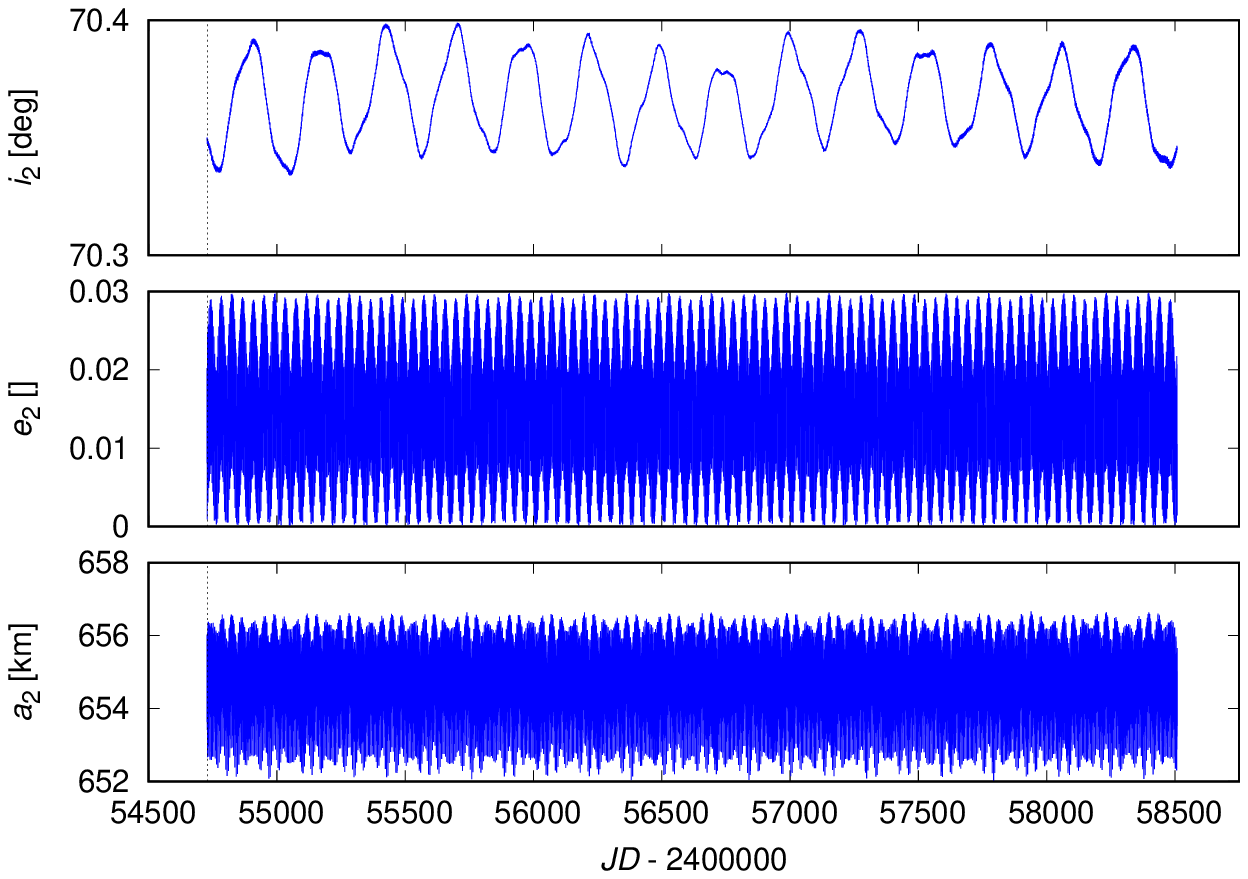}
\caption{
Evolution of the osculating elements over a time span of 3780\,d,
shown for the semimajor axes~$a_1$, $a_2$,
eccentricities~$e_1$, $e_2$,
and inclinations~$i_1$, $i_2$.
Oscillations are mostly caused by the multipoles of Kleopatra.
The moons mutually interact only weakly.
}
\label{216_fitting33_360DEG_orbit1}
\end{figure*}

Let us finally present the best-fit model, with
$\chi^2 = \chi^2_{\rm sky} + \chi^2_{\rm sky2} + 0.3\,\chi^2_{\rm ao} = 368$.
Its parameters are summarized in Tab.~\ref{tab3}
and the results in
Figs.~\ref{216_fitting33_360DEG_chi2_SKY_uv},
\ref{216_fitting33_360DEG_chi2_SKY2_uv},
\ref{216_fitting33_360DEG_chi2_AO}.
The orbits can be perhaps seen more clearly
if we plot the three datasets separately
(Figs.~\ref{216_fitting33_360DEG_DESCAMPS_chi2_SKY_uv},
\ref{216_fitting33_360DEG_DESCAMPS_chi2_SKY2_uv}).
To emphasize the orbital elements are not constants in our dynamical model,
we demonstrate it in accompanying
Fig.~\ref{216_fitting33_360DEG_orbit1}.
The oscillations of $a$, $e$, $i$ for the inner moon reach
$6\,{\rm km}$, $0.04$, $0.5^\circ$, respectively.
The inclinations with respect to Kleopatra's equator are close to zero.
The dominant short-periodic terms are directly related
to the ${\sim}\,5.4$-hour rotation of Kleopatra.
The longer 100 and 270-day periods of inclinations
correspond to the nodal precession
if the reference plane is the equator.

The RMS residua of absolute astrometric measurements are approximately
$17\,{\rm mas}$ (or $23\,{\rm mas}$ for relative),
which should be compared to the assumed uncertainties of
$10\,{\rm mas}$.
This fit is acceptable,
with the reduced $\chi^2_{\rm R} = 1.71$ (or $2.35$),
especially because we do not see significant systematic problems.
The values may be increased due to
underestimated uncertainties of astrometric observations,
remaining systematics related to the tangential (along-track) motion,
the shape model ($C_{\ell m}$, $S_{\ell m}$) is not correct,
and/or the density distribution is not uniform.

As there is no unique solution, we also present an alternative model,
namely with $\chi^2 = 381$ (Tab.~\ref{tab3}, right).
It has a slightly higher mass~$m_1$ (by 10\,\%),
and adjusted periods $P_1$, $P_2$,
so that the number of revolutions over $t_2-t_1$
remains the same, with epochs $E_1 = 2149.08$, $E_2 = 1407.55$.
On the other hand, the moons' masses $m_2$, $m_3$
are substantially higher (by a~factor of 2 to 3).
Last but not least, 
we can use the difference between these models to estimate
realistic uncertainties of the parameters.


\begin{table}
\caption{Multipole coefficients of Kleopatra's gravitational field,
using the ADAM model and constant density.
The normalisation is given by Eq.~(\ref{eq:U}).
The reference radius is $R = 59.633\,{\rm km}$.
}
\label{tab1}
\centering
\tiny
\begin{tabular}{lrlr}
$C_{00}$ & $1.00000000\phantom{\cdot10^{-00}}$ & \\
$C_{10}$ & $0.00000000\phantom{\cdot10^{-00}}$ & \\
$C_{11}$ & $0.00000000\phantom{\cdot10^{-00}}$ & $S_{11}$ & $0.00000000\phantom{\cdot10^{-00}}$\\
$C_{20}$ & $-7.65106929\cdot10^{-1}$ & \\
$C_{21}$ & $3.98110264\cdot10^{-4}$ & $S_{21}$ & $-3.07876838\cdot10^{-4}$\\
$C_{22}$ & $3.59335850\cdot10^{-1}$ & $S_{22}$ & $-8.65906339\cdot10^{-5}$\\
$C_{30}$ & $1.49466956\cdot10^{-2}$ & \\
$C_{31}$ & $-5.24916471\cdot10^{-2}$ & $S_{31}$ & $5.20018496\cdot10^{-4}$\\
$C_{32}$ & $-1.48712568\cdot10^{-3}$ & $S_{32}$ & $2.52505000\cdot10^{-3}$\\
$C_{33}$ & $1.17882333\cdot10^{-2}$ & $S_{33}$ & $-3.43079734\cdot10^{-4}$\\
$C_{40}$ & $1.30914835\phantom{\cdot10^{-00}}$ & \\
$C_{41}$ & $-1.41497526\cdot10^{-2}$ & $S_{41}$ & $7.18145896\cdot10^{-4}$\\
$C_{42}$ & $-1.39568658\cdot10^{-1}$ & $S_{42}$ & $-1.72827301\cdot10^{-3}$\\
$C_{43}$ & $3.44681126\cdot10^{-4}$ & $S_{43}$ & $-6.95352555\cdot10^{-5}$\\
$C_{44}$ & $1.53908741\cdot10^{-2}$ & $S_{44}$ & $5.76718751\cdot10^{-4}$\\
$C_{50}$ & $-3.01286209\cdot10^{-2}$ & \\
$C_{51}$ & $1.39623684\cdot10^{-1}$ & $S_{51}$ & $-4.39146849\cdot10^{-5}$\\
$C_{52}$ & $9.49158788\cdot10^{-4}$ & $S_{52}$ & $-3.50208422\cdot10^{-3}$\\
$C_{53}$ & $-6.31945029\cdot10^{-3}$ & $S_{53}$ & $1.44844074\cdot10^{-6}$\\
$C_{54}$ & $7.63010533\cdot10^{-5}$ & $S_{54}$ & $2.87991860\cdot10^{-4}$\\
$C_{55}$ & $7.06444516\cdot10^{-4}$ & $S_{55}$ & $9.22871681\cdot10^{-6}$\\
$C_{60}$ & $-2.92621603\phantom{\cdot10^{-00}}$ & \\
$C_{61}$ & $5.01230966\cdot10^{-2}$ & $S_{61}$ & $-3.30583966\cdot10^{-3}$\\
$C_{62}$ & $1.41764499\cdot10^{-1}$ & $S_{62}$ & $2.29130383\cdot10^{-3}$\\
$C_{63}$ & $-9.12143591\cdot10^{-4}$ & $S_{63}$ & $2.24108452\cdot10^{-4}$\\
$C_{64}$ & $-4.29428304\cdot10^{-3}$ & $S_{64}$ & $-1.93439069\cdot10^{-4}$\\
$C_{65}$ & $1.33372930\cdot10^{-5}$ & $S_{65}$ & $-9.83952170\cdot10^{-6}$\\
$C_{66}$ & $3.04072429\cdot10^{-4}$ & $S_{66}$ & $3.14393964\cdot10^{-5}$\\
$C_{70}$ & $-1.49050705\cdot10^{-3}$ & \\
$C_{71}$ & $-4.94081000\cdot10^{-1}$ & $S_{71}$ & $-2.54792669\cdot10^{-3}$\\
$C_{72}$ & $1.49282198\cdot10^{-3}$ & $S_{72}$ & $6.59185873\cdot10^{-3}$\\
$C_{73}$ & $7.58761100\cdot10^{-3}$ & $S_{73}$ & $1.54865702\cdot10^{-5}$\\
$C_{74}$ & $-1.06667828\cdot10^{-4}$ & $S_{74}$ & $-2.55582680\cdot10^{-4}$\\
$C_{75}$ & $-2.15423464\cdot10^{-4}$ & $S_{75}$ & $-4.16149717\cdot10^{-6}$\\
$C_{76}$ & $5.72095667\cdot10^{-6}$ & $S_{76}$ & $1.01143177\cdot10^{-5}$\\
$C_{77}$ & $1.54723186\cdot10^{-5}$ & $S_{77}$ & $1.12688336\cdot10^{-6}$\\
$C_{80}$ & $7.61525254\phantom{\cdot10^{-00}}$ & \\
$C_{81}$ & $-1.66415917\cdot10^{-1}$ & $S_{81}$ & $1.15002782\cdot10^{-2}$\\
$C_{82}$ & $-2.12136010\cdot10^{-1}$ & $S_{82}$ & $-4.07662706\cdot10^{-3}$\\
$C_{83}$ & $2.01880918\cdot10^{-3}$ & $S_{83}$ & $-4.60678358\cdot10^{-4}$\\
$C_{84}$ & $3.27454389\cdot10^{-3}$ & $S_{84}$ & $1.62346625\cdot10^{-4}$\\
$C_{85}$ & $-2.40746340\cdot10^{-5}$ & $S_{85}$ & $1.09881420\cdot10^{-5}$\\
$C_{86}$ & $-6.84753900\cdot10^{-5}$ & $S_{86}$ & $-7.21260224\cdot10^{-6}$\\
$C_{87}$ & $3.52331279\cdot10^{-7}$ & $S_{87}$ & $-2.79130886\cdot10^{-7}$\\
$C_{88}$ & $3.55651688\cdot10^{-6}$ & $S_{88}$ & $7.47645373\cdot10^{-7}$\\
$C_{90}$ & $4.18361848\cdot10^{-1}$ & \\
$C_{91}$ & $1.06717720\phantom{\cdot10^{-00}}$ & $S_{91}$ & $-5.47346878\cdot10^{-4}$\\
$C_{92}$ & $-1.15403753\cdot10^{-2}$ & $S_{92}$ & $-1.39184073\cdot10^{-2}$\\
$C_{93}$ & $-2.74503944\cdot10^{-2}$ & $S_{93}$ & $-1.53156145\cdot10^{-3}$\\
$C_{94}$ & $2.12636564\cdot10^{-4}$ & $S_{94}$ & $3.22499931\cdot10^{-4}$\\
$C_{95}$ & $1.74908520\cdot10^{-4}$ & $S_{95}$ & $3.27374899\cdot10^{-6}$\\
$C_{96}$ & $-4.45393947\cdot10^{-6}$ & $S_{96}$ & $-6.47023483\cdot10^{-6}$\\
$C_{97}$ & $-3.52023978\cdot10^{-6}$ & $S_{97}$ & $-2.27651271\cdot10^{-7}$\\
$C_{98}$ & $1.20584507\cdot10^{-7}$ & $S_{98}$ & $1.64507039\cdot10^{-7}$\\
$C_{99}$ & $1.83429337\cdot10^{-7}$ & $S_{99}$ & $3.05515155\cdot10^{-8}$\\
$C_{10,0}$ & $-2.21145150\cdot10^{1}\phantom{^+}$ & \\
$C_{10,1}$ & $5.03865729\cdot10^{-1}$ & $S_{10,1}$ & $-3.33850136\cdot10^{-2}$\\
$C_{10,2}$ & $4.00901809\cdot10^{-1}$ & $S_{10,2}$ & $9.98417914\cdot10^{-3}$\\
$C_{10,3}$ & $-4.78661768\cdot10^{-3}$ & $S_{10,3}$ & $9.35952923\cdot10^{-4}$\\
$C_{10,4}$ & $-3.83595725\cdot10^{-3}$ & $S_{10,4}$ & $-2.25073571\cdot10^{-4}$\\
$C_{10,5}$ & $4.13709990\cdot10^{-5}$ & $S_{10,5}$ & $-1.37689312\cdot10^{-5}$\\
$C_{10,6}$ & $4.30269516\cdot10^{-5}$ & $S_{10,6}$ & $4.84072887\cdot10^{-6}$\\
$C_{10,7}$ & $-4.17610659\cdot10^{-7}$ & $S_{10,7}$ & $1.86498646\cdot10^{-7}$\\
$C_{10,8}$ & $-6.83864198\cdot10^{-7}$ & $S_{10,8}$ & $-1.39471718\cdot10^{-7}$\\
$C_{10,9}$ & $6.37173159\cdot10^{-9}$ & $S_{10,9}$ & $-2.88792150\cdot10^{-9}$\\
$C_{10,10}$ & $2.80465119\cdot10^{-8}$ & $S_{10,10}$ & $1.02591965\cdot10^{-8}$\\
\end{tabular}
\end{table}

\begin{table*}
\caption{Convergence test of the multipole approximation.
The acceleration components $\vec a = (a_x, a_y, a_z)$ were evaluated
for the position vector $\vec r = (500\,{\rm km}; 0; 0).$}
\label{tab2}
\centering
\tiny
\begin{tabular}{rrrl}
\multicolumn{1}{c}{$a_x\ [{\rm m}\,{\rm s}^{-2}]$} &
\multicolumn{1}{c}{$a_y\ [{\rm m}\,{\rm s}^{-2}]$} &
\multicolumn{1}{c}{$a_z\ [{\rm m}\,{\rm s}^{-2}]$} &
description \\
\hline
\vrule height8pt width0pt
$-1.23875008\cdot10^{-3}$ & $0.00000000\phantom{{}\cdot10^{-0}}$ & $0.00000000\phantom{{}\cdot10^{-0}}$ & point mass\\
$-1.32251141\cdot10^{-3}$ & $3.52412309\cdot10^{-8}$ & $-3.27218005\cdot10^{-8}$ & brute force\\
$-1.23875008\cdot10^{-3}$ & $0.00000000\phantom{{}\cdot10^{-0}}$ & $0.00000000\phantom{{}\cdot10^{-0}}$ & multipole, 0\\
$-1.23875008\cdot10^{-3}$ & $0.00000000\phantom{{}\cdot10^{-0}}$ & $0.00000000\phantom{{}\cdot10^{-0}}$ & multipole, 1\\
$-1.31595722\cdot10^{-3}$ & $-9.15458551\cdot10^{-9}$ & $2.10446228\cdot10^{-8}$ & multipole, 2\\
$-1.31810548\cdot10^{-3}$ & $-4.32382838\cdot10^{-8}$ & $-7.29497823\cdot10^{-8}$ & multipole, 3\\
$-1.32205774\cdot10^{-3}$ & $2.39696605\cdot10^{-8}$ & $-3.72801209\cdot10^{-8}$ & multipole, 4\\
$-1.32228394\cdot10^{-3}$ & $2.52638733\cdot10^{-8}$ & $-3.83029760\cdot10^{-8}$ & multipole, 5\\
$-1.32248271\cdot10^{-3}$ & $3.37726230\cdot10^{-8}$ & $-3.39267296\cdot10^{-8}$ & multipole, 6\\
$-1.32250036\cdot10^{-3}$ & $3.42765669\cdot10^{-8}$ & $-3.32858996\cdot10^{-8}$ & multipole, 7\\
$-1.32251056\cdot10^{-3}$ & $3.50906194\cdot10^{-8}$ & $-3.28680108\cdot10^{-8}$ & multipole, 8\\
$-1.32251185\cdot10^{-3}$ & $3.51653783\cdot10^{-8}$ & $-3.27725086\cdot10^{-8}$ & multipole, 9\\
$-1.32251239\cdot10^{-3}$ & $3.52352260\cdot10^{-8}$ & $-3.27371874\cdot10^{-8}$ & multipole, 10\\
\end{tabular}
\end{table*}

\begin{table*}
\caption{Best-fit (left) and alternative (middle) model parameters,
together with realistic uncertainties (right).
Orbital elements of the moons are osculating,
for the epoch $T_0 = 2454728.761806$
(cf.~Fig.~\ref{216_fitting33_360DEG_orbit1}).}
\label{tab3}
\centering
\begin{tabular}{lrrlr}
var. & val. & val. & unit & $\sigma$ \\
\hline
\vrule height10pt width0pt
$m_1         $ & $   1.492735\cdot10^{-12}   $ & $   1.651829\cdot10^{-12}  $ & $M_{\rm S}$ & $ 0.16\cdot10^{-12} $ \\
$m_2         $ & $   2\cdot10^{-16}          $ & $   4\cdot10^{-16}         $ & $M_{\rm S}$ & $ 2\cdot10^{-16}    $ \\
$m_3         $ & $   3\cdot10^{-16}          $ & $   9\cdot10^{-16}         $ & $M_{\rm S}$ & $ 3\cdot10^{-16}    $ \\
$P_1         $ & $   1.822359                $ & $   1.818203               $ & day         & $ 0.004156          $ \\
$\log e_1    $ & $  -3.991                   $ & $  -4.100                  $ & 1           & $ -3$ (i.e. 0.001)    \\
$i_1         $ & $  70.104                   $ & $  68.719                  $ & deg         & $ 1.0               $ \\
$\Omega_1    $ & $ 252.920                   $ & $ 253.751                  $ & deg         & $ 1.0               $ \\
$\varpi_1    $ & $   0.089                   $ & $  13.892                  $ & deg         & $ 10.0              $ \\
$\lambda_1   $ & $  59.665                   $ & $  60.565                  $ & deg         & $ 1.0               $ \\
$P_2         $ & $   2.745820                $ & $   2.740999               $ & day         & $ 0.004820          $ \\
$\log e_2    $ & $  -3.998                   $ & $  -4.138                  $ & 1           & $ -3                $ \\
$i_2         $ & $  70.347                   $ & $  69.383                  $ & deg         & $ 1.0               $ \\
$\Omega_2    $ & $ 252.954                   $ & $ 252.033                  $ & deg         & $ 1.0               $ \\
$\varpi_2    $ & $   1.601                   $ & $  -9.757                  $ & deg         & $ 10.0              $ \\
$\lambda_2   $ & $ 108.357                   $ & $ 107.865                  $ & deg         & $ 1.0               $ \\
$l_{\rm pole}$ & $  72.961                   $ & $  73.472                  $ & deg         & $ 1.0               $ \\
$b_{\rm pole}$ & $  19.628                   $ & $  20.480                  $ & deg         & $ 1.0               $ \\
\hline
$n_{\rm sky}         $ & $  66  $ & $  66$ \\
$n_{\rm sky2}        $ & $  28  $ & $  28$ \\
$n_{\rm ao}          $ & $3240  $ & $3240$ \\
\hline
$\chi^2_{\rm sky}    $ & $ 113  $ & $ 124$ \\
$\chi^2_{\rm sky2}   $ & $  66  $ & $  78$ \\
$\chi^2_{\rm ao}     $ & $ 621  $ & $ 584$ \\
$\chi^2              $ & $ 368  $ & $ 381$ \\
\hline
$\chi^2_{\rm R\,sky} $ & $ 1.71 $ & $1.87$ \\
$\chi^2_{\rm R\,sky2}$ & $ 2.35 $ & $2.78$ \\
$\chi^2_{\rm R\,ao}  $ & $ 0.19 $ & $0.18$ \\
\end{tabular}
\tablefoot{
$m_1$ denotes the mass of body~1 (i.e.~Kleopatra),
$m_2$ body~2 (1st moon),
$m_3$ body~3 (2nd moon),
$P_1$ the orbital period of the 1st orbit,
$e_1$ eccentricity,
$i_1$ inclination,
$\Omega_1$ longitude of node,
$\varpi_1$ longitude of pericentre,
$\lambda_1$ true longitude,
etc. of the 2nd orbit;
$l_{\rm pole}$ ecliptic longitude of Kleopatra's rotation pole,
$b_{\rm pole}$ ecliptic latitude;
$n$~numbers of observations (SKY, SKY2, AO),
$\chi^2$~values,
$\chi^2_{\rm R} \equiv \chi^2/n$ reduced values.
The angular orbital elements are expressed in the standard stellar reference frame.
If the orbits lie in the equatorial plane of body~1, they fulfil
$i = 90^\circ-b_{\rm pole}$,
$\Omega = 180^\circ+l_{\rm pole}$.
}
\end{table*}



\section{Implications for the moons}

The nominal periods of the moons,
$P_1 = 1.822359\,{\rm d}$,
$P_2 = 2.745820\,{\rm d}$,
--- or the semimajor axes 499 and 655\,km ---
are relatively close to each other.
In our nominal model, the mutual interactions are weak,
but if we would artificially increase the masses,
they soon become strong.
The upper limit for the stability of the moon system is about
$m_2, m_3 \simeq 3\cdot10^{-15}\,M_\odot$.
Eccentricities hardly can be larger than $e_1, e_2 \simeq 0.1$,
because orbits then start to perturb and cross each other.
Such closely-packed moon system strongly indicates a common origin.

Moreover, the period ratio is close to the 3:2 mean-motion resonance,
with $P_2/P_1 \doteq 1.507$ (cf.~Fig.~\ref{216_fitting9_BIGGRID_P1_P2_min_32}).
We should specify the resonant condition more precisely though,
because the perihelion precession rate~$\dot\varpi$
is non-negligible in the vicinity of an oblate body
(namely, $n_1 = 197\,{\rm deg}\,{\rm d}^{-1}$, $\dot\varpi_1 \simeq 3\,{\rm deg}\,{\rm d}^{-1}$).
The resonant angle is defined as:
\begin{equation}
\sigma = 3\lambda_2 - 2\lambda_1 - \varpi_1\,,
\end{equation}
or alternatively $\varpi_2$ instead of $\varpi_1$.
The stable configuration is expected when conjunctions occur
in the apocentre of the outer moon (or the pericentre of the inner moon).
On the other hand, it's not a circular restricted three-body problem:
(i)~the moons have comparable masses,
(ii)~the central body is irregular which induces perturbations
on the synodic rotation time scale (sideric $P = 0.224386\,{\rm d}$).
According to our tests with bodies purposely placed in the exact resonance,
or offset in the longitude so that the libration amplitude is ${\sim}\,90^\circ$,
regular librations are notable only if the initial (osculating) eccentricities
$e_1, e_2 \gtrsim 10^{-2}$ (cf.~Fig.~\ref{216_fitting33_360DEG_orbit1}).
In the current best-fit configuration, they ain't.

In the future, it is important to better constrain the masses of moons.
This task would require an extended astrometric dataset
compared to what is available at the moment.
If their low densities are confirmed, the interpretation would be
that regolith making up both Kleopatra
and the moons is relatively 'fine'
(with block sizes smaller than the moon diameters)
and it is more compressed in Kleopatra and less compressed in the moons.
On contrary, if densities are high the interpretation would be the opposite:
'coarse' regolith in Kleopatra and monolithic material in the moons.
This does not seem so likely, though. 

For comparison, let us recall basic parameters of the Haumea moon system
\citep{Ortiz_2017Natur.550..219O,Dunham_2019ApJ...877...41D}.
Although everything is about 10 times larger,
the central body is very elongated triaxial ellipsoid (2.0:1.6:1),
which is rapidly rotating (3,9\,h).
The closest to the centre is the ring system,
with ring particles orbiting close to the 3:1 spin-orbit resonance.
There are two moons, inner Namaka and outer Hi'iaka,
which are close to the 8:3 mean-motion resonance.
The inner orbit is inclined, possibly perturbed by the ellipsoidal body,
the outer is co-planar with the equator and the ring.
A distinct collisional family related to Haumea
was also identified \citep{Brown_2007Natur.446..294B,Leinhardt_2010ApJ...714.1789L}.

Clearly, the Kleopatra moon system is somewhat different
--- its moons are co-planar and more closely packed.
There is no ring and no family \citep{Nesvorny_etal_2015aste.book..297N}.
Nevertheless, the nearly-critical rotation as well as
the mass ratios of the order of $10^{-3}$ vs. $10^{-4}$ are similar.
Consequently, moon formation by mass shedding,
after a rotational fission initiated by a low-energy impact
(as in \citealt{Ortiz_2012MNRAS.419.2315O}) seems viable.


\section{Conclusions}

Having revised the mass of (216) Kleopatra,
it is worth revising the interpretation of its shape
(see the paper by Marchis et al.).
We plan to use our multipole model
also for analyses of other triple systems observed by the VLT/SPHERE
(e.g., (45) Eugenia, (130) Elektra).

In this paper, we focus on future improvements of dynamical models.
According to our preliminary tests, it should be possible
to measure also angular velocities,
because astrometric positions measured on close-in-time images
are aligned with derived orbits.
Even if the velocity magnitude is not correct,
because of residual seeing and an under-corrected PSF,
it is sufficient to measure its direction (`sign'),
which would prevent some of the ambiguities.

In our current model, we assume a fixed shape
(derived by other methods).
During the fitting, we let the pole orientation to vary slightly,
although the shape and pole are always correlated.
Moreover, we only fit silhouettes, which is surely inferior
(compared to other methods).
While it is not easy for us to combine a full N-body modelling
with a full shape modelling,
it may be viable to treat the multipole coefficients $C_{\ell m}$, $S_{\!\ell m}$
as free parameters.
If adaptive-optics observations of asteroid moon systems
will continue in the future,
we may be at the dawn of asteroid `geodesy' from the ground.






\begin{acknowledgements}
We thank an anonymous referee for valuable comments.
This work has been supported by the Czech Science Foundation through grant
21-11058S (M.~Bro\v z, D.~Vokrouhlick\'y),
20-08218S (J.~Hanu\v s, J.~\v Durech),
and by the Charles University Research program No. UNCE/SCI/023.
This material is partially based upon work supported by the National Science Foundation under Grant No. 1743015.
P.~Vernazza, A.~Drouard, M.~Ferrais and B.~Carry were supported by CNRS/INSU/PNP.
M.M. was supported by the National Aeronautics and Space Administration under grant No. 80NSSC18K0849 issued through the Planetary Astronomy Program.
The work of TSR was carried out through grant APOSTD/2019/046 by Generalitat Valenciana (Spain).
This work was supported by the MINECO (Spanish Ministry of Economy) through grant RTI2018-095076-B-C21 (MINECO/FEDER, UE).
The research leading to these results has received funding from the ARC grant for Concerted Research Actions, financed by the Wallonia-Brussels Federation.
TRAPPIST is a project funded by the Belgian Fonds (National) de la Recherche Scientifique (F.R.S.-FNRS) under grant FRFC 2.5.594.09.F.
TRAPPIST-North is a project funded by the University of Liège, and performed in collaboration with Cadi Ayyad University of Marrakesh.
E. Jehin is a FNRS Senior Research Associate.
The data presented herein were obtained partially at the W. M. Keck Observatory, which is operated as a scientific partnership among the California Institute of Technology, the University of California and the National Aeronautics and Space Administration. The Observatory was made possible by the generous financial support of the W. M. Keck Foundation. The authors wish to recognize and acknowledge the very significant cultural role and reverence that the summit of Maunakea has always had within the indigenous Hawaiian community.  We are most fortunate to have the opportunity to conduct observations from this mountain.
\end{acknowledgements}

\bibliographystyle{aa}
\bibliography{references}

\begin{thebibliography}{20}
\expandafter\ifx\csname natexlab\endcsname\relax\def\natexlab#1{#1}\fi

\bibitem[{{Bertotti} {et~al.}(2003){Bertotti}, {Farinella}, \&
  {Vokrouhlick{\'y}}}]{Bertotti_etal_2003ASSL..293.....B}
{Bertotti}, B., {Farinella}, P., \& {Vokrouhlick{\'y}}, D. 2003, {Physics of
  the Solar System --- Dynamics and Evolution, Space Physics, and Spacetime
  Structure.}, Vol. 293 (Kluwer)

\bibitem[{{Beuzit} {et~al.}(2019){Beuzit}, {Vigan}, {Mouillet}, {Dohlen},
  {Gratton}, {Boccaletti}, {Sauvage}, {Schmid}, {Langlois}, {Petit},
  {Baruffolo}, {Feldt}, {Milli}, {Wahhaj}, {Abe}, {Anselmi}, {Antichi},
  {Barette}, {Baudrand}, {Baudoz}, {Bazzon}, {Bernardi}, {Blanchard}, {Brast},
  {Bruno}, {Buey}, {Carbillet}, {Carle}, {Cascone}, {Chapron}, {Charton},
  {Chauvin}, {Claudi}, {Costille}, {De Caprio}, {de Boer}, {Delboulb{\'e}},
  {Desidera}, {Dominik}, {Downing}, {Dupuis}, {Fabron}, {Fantinel}, {Farisato},
  {Feautrier}, {Fedrigo}, {Fusco}, {Gigan}, {Ginski}, {Girard}, {Giro},
  {Gisler}, {Gluck}, {Gry}, {Henning}, {Hubin}, {Hugot}, {Incorvaia}, {Jaquet},
  {Kasper}, {Lagadec}, {Lagrange}, {Le Coroller}, {Le Mignant}, {Le Ruyet},
  {Lessio}, {Lizon}, {Llored}, {Lundin}, {Madec}, {Magnard}, {Marteaud},
  {Martinez}, {Maurel}, {M{\'e}nard}, {Mesa}, {M{\"o}ller-Nilsson}, {Moulin},
  {Moutou}, {Orign{\'e}}, {Parisot}, {Pavlov}, {Perret}, {Pragt}, {Puget},
  {Rabou}, {Ramos}, {Reess}, {Rigal}, {Rochat}, {Roelfsema}, {Rousset}, {Roux},
  {Saisse}, {Salasnich}, {Santambrogio}, {Scuderi}, {Segransan}, {Sevin},
  {Siebenmorgen}, {Soenke}, {Stadler}, {Suarez}, {Tiph{\`e}ne}, {Turatto},
  {Udry}, {Vakili}, {Waters}, {Weber}, {Wildi}, {Zins}, \&
  {Zurlo}}]{Beuzit_2019A&A...631A.155B}
{Beuzit}, J.~L., {Vigan}, A., {Mouillet}, D., {et~al.} 2019, \aap, 631, A155

\bibitem[{{Bro{\v{z}}}(2017)}]{Broz_2017ApJS..230...19B}
{Bro{\v{z}}}, M. 2017, \apjs, 230, 19

\bibitem[{{Brown} {et~al.}(2007){Brown}, {Barkume}, {Ragozzine}, \&
  {Schaller}}]{Brown_2007Natur.446..294B}
{Brown}, M.~E., {Barkume}, K.~M., {Ragozzine}, D., \& {Schaller}, E.~L. 2007,
  \nat, 446, 294

\bibitem[{{Bur\v{s}a} {et~al.}(1993){Bur\v{s}a}, {Karsk\'{y}}, \&
  {Kosteleck\'{y}}}]{Bursa_etal_1993}
{Bur\v{s}a}, M., {Karsk\'{y}}, G., \& {Kosteleck\'{y}}, J. 1993, {Dynamika
  um\v{e}l\'{y}ch dru\v{z}ic v t\'{i}hov\'{e}m poli Zem\v{e}.} (Academia)

\bibitem[{{Descamps} {et~al.}(2011){Descamps}, {Marchis}, {Berthier}, {Emery},
  {Duch{\^e}ne}, {de Pater}, {Wong}, {Lim}, {Hammel}, {Vachier}, {Wiggins},
  {Teng-Chuen-Yu}, {Peyrot}, {Pollock}, {Assafin}, {Vieira-Martins}, {Camargo},
  {Braga-Ribas}, \& {Macomber}}]{Descamps_etal_2011Icar..211.1022D}
{Descamps}, P., {Marchis}, F., {Berthier}, J., {et~al.} 2011, \icarus, 211,
  1022

\bibitem[{{Dunham} {et~al.}(2019){Dunham}, {Desch}, \&
  {Probst}}]{Dunham_2019ApJ...877...41D}
{Dunham}, E.~T., {Desch}, S.~J., \& {Probst}, L. 2019, \apj, 877, 41

\bibitem[{{Giorgini} {et~al.}(1996){Giorgini}, {Yeomans}, {Chamberlin},
  {Chodas}, {Jacobson}, {Keesey}, {Lieske}, {Ostro}, {Standish}, \&
  {Wimberly}}]{Giorgini_etal_1996DPS....28.2504G}
{Giorgini}, J.~D., {Yeomans}, D.~K., {Chamberlin}, A.~B., {et~al.} 1996, in
  AAS/Division for Planetary Sciences Meeting Abstracts, Vol.~28, AAS/Division
  for Planetary Sciences Meeting Abstracts \#28, 25.04

\bibitem[{{Goldreich}(1965)}]{Goldreich_1965AJ.....70....5G}
{Goldreich}, P. 1965, \aj, 70, 5

\bibitem[{{Leinhardt} {et~al.}(2010){Leinhardt}, {Marcus}, \&
  {Stewart}}]{Leinhardt_2010ApJ...714.1789L}
{Leinhardt}, Z.~M., {Marcus}, R.~A., \& {Stewart}, S.~T. 2010, \apj, 714, 1789

\bibitem[{{Levison} \& {Duncan}(1994)}]{Levison_Duncan_1994Icar..108...18L}
{Levison}, H.~F. \& {Duncan}, M.~J. 1994, \icarus, 108, 18

\bibitem[{Nelder \& Mead(1965)}]{Nelder_Mead_1965}
Nelder, J.~A. \& Mead, R. 1965, The Computer Journal, 7, 308

\bibitem[{{Nemravov{\'a}} {et~al.}(2016){Nemravov{\'a}}, {Harmanec},
  {Bro{\v{z}}}, {Vokrouhlick{\'y}}, {Mourard}, {Hummel}, {Cameron}, {Matthews},
  {Bolton}, {Bo{\v{z}}i{\'c}}, {Chini}, {Dembsky}, {Engle}, {Farrington},
  {Grunhut}, {Guenther}, {Guinan}, {Kor{\v{c}}{\'a}kov{\'a}}, {Koubsk{\'y}},
  {K{\v{r}}{\'\i}{\v{c}}ek}, {Kuschnig}, {Mayer}, {McCook}, {Moffat},
  {Nardetto}, {Pr{\v{s}}a}, {Ribeiro}, {Rowe}, {Rucinski}, {{\v{S}}koda},
  {{\v{S}}lechta}, {Tallon-Bosc}, {Votruba}, {Weiss}, {Wolf}, {Zasche}, \&
  {Zavala}}]{Nemravova_etal_2016A&A...594A..55N}
{Nemravov{\'a}}, J.~A., {Harmanec}, P., {Bro{\v{z}}}, M., {et~al.} 2016, \aap,
  594, A55

\bibitem[{{Nesvorn{\'y}} {et~al.}(2015){Nesvorn{\'y}}, {Bro{\v z}}, \&
  {Carruba}}]{Nesvorny_etal_2015aste.book..297N}
{Nesvorn{\'y}}, D., {Bro{\v z}}, M., \& {Carruba}, V. 2015, {Identification and
  Dynamical Properties of Asteroid Families}, ed. P.~{Michel}, F.~E. {DeMeo},
  \& W.~F. {Bottke} (Univ. Arizona Press), 297--321

\bibitem[{{Ortiz} {et~al.}(2017){Ortiz}, {Santos-Sanz}, {Sicardy},
  {Benedetti-Rossi}, {B{\'e}rard}, {Morales}, {Duffard}, {Braga-Ribas}, {Hopp},
  {Ries}, {Nascimbeni}, {Marzari}, {Granata}, {P{\'a}l}, {Kiss}, {Pribulla},
  {Kom{\v{z}}{\'\i}k}, {Hornoch}, {Pravec}, {Bacci}, {Maestripieri}, {Nerli},
  {Mazzei}, {Bachini}, {Martinelli}, {Succi}, {Ciabattari}, {Mikuz},
  {Carbognani}, {Gaehrken}, {Mottola}, {Hellmich}, {Rommel},
  {Fern{\'a}ndez-Valenzuela}, {Campo Bagatin}, {Cikota}, {Cikota}, {Lecacheux},
  {Vieira-Martins}, {Camargo}, {Assafin}, {Colas}, {Behrend}, {Desmars},
  {Meza}, {Alvarez-Candal}, {Beisker}, {Gomes-Junior}, {Morgado}, {Roques},
  {Vachier}, {Berthier}, {Mueller}, {Madiedo}, {Unsalan}, {Sonbas}, {Karaman},
  {Erece}, {Koseoglu}, {Ozisik}, {Kalkan}, {Guney}, {Niaei}, {Satir},
  {Yesilyaprak}, {Puskullu}, {Kabas}, {Demircan}, {Alikakos}, {Charmandaris},
  {Leto}, {Ohlert}, {Christille}, {Szak{\'a}ts}, {Tak{\'a}csn{\'e} Farkas},
  {Varga-Vereb{\'e}lyi}, {Marton}, {Marciniak}, {Bartczak}, {Santana-Ros},
  {Butkiewicz-B{\k{a}}k}, {Dudzi{\'n}ski}, {Al{\'\i}-Lagoa}, {Gazeas},
  {Tzouganatos}, {Paschalis}, {Tsamis}, {S{\'a}nchez-Lavega},
  {P{\'e}rez-Hoyos}, {Hueso}, {Guirado}, {Peris}, \&
  {Iglesias-Marzoa}}]{Ortiz_2017Natur.550..219O}
{Ortiz}, J.~L., {Santos-Sanz}, P., {Sicardy}, B., {et~al.} 2017, \nat, 550, 219

\bibitem[{{Ortiz} {et~al.}(2012){Ortiz}, {Thirouin}, {Campo Bagatin},
  {Duffard}, {Licandro}, {Richardson}, {Santos-Sanz}, {Morales}, \&
  {Benavidez}}]{Ortiz_2012MNRAS.419.2315O}
{Ortiz}, J.~L., {Thirouin}, A., {Campo Bagatin}, A., {et~al.} 2012, \mnras,
  419, 2315

\bibitem[{{Ostro} {et~al.}(2000){Ostro}, {Hudson}, {Nolan}, {Margot},
  {Scheeres}, {Campbell}, {Magri}, {Giorgini}, \& {Yeomans}}]{Ostro2000}
{Ostro}, S.~J., {Hudson}, R.~S., {Nolan}, M.~C., {et~al.} 2000, Science, 288,
  836

\bibitem[{{Palisa}(1880)}]{Palisa_1880AN.....98..129P}
{Palisa}, J. 1880, Astronomische Nachrichten, 98, 129

\bibitem[{{Shepard} {et~al.}(2018){Shepard}, {Timerson}, {Scheeres}, {Benner},
  {Giorgini}, {Howell}, {Magri}, {Nolan}, {Springmann}, {Taylor}, \&
  {Virkki}}]{Shepard_etal_2018Icar..311..197S}
{Shepard}, M.~K., {Timerson}, B., {Scheeres}, D.~J., {et~al.} 2018, \icarus,
  311, 197

\bibitem[{{Si}(2006)}]{Si_2006}
{Si}, H. 2006, Available at \url{http://wias-berlin.de/software/tetgen/}

\end{thebibliography}

\end{document}